\documentclass[aps,prd, showpacs, amsmath,amssymb,eqsecnum]{revtex4}
\usepackage{graphicx}%
\usepackage{subfigure}
\usepackage{mathptmx}
\usepackage{dcolumn}
\usepackage{bm}
\usepackage{tabls}
\newcommand{\beq}{\begin{equation}}
\newcommand{\eeq}{\end{equation}}
\newcommand{\be}{\begin{eqnarray}}
\newcommand{\ee}{\end{eqnarray}}
\long\def\hidestart#1\hideend{}
\begin{document}

\title{Investigation of Lattice QCD with Wilson fermions with Gaussian 
Smearing}

\author{Asit K. De}
\email{asitk.de@saha.ac.in}

\author{A. Harindranath}
\email{a.harindranath@saha.ac.in}

\author{Jyotirmoy Maiti}
\email{jyotirmoy.maiti@saha.ac.in}

\affiliation{Theory Group, Saha Institute of Nuclear Physics \\
 1/AF Bidhan Nagar, Kolkata 700064, India}

\date{December 28, 2007}

\begin{abstract}

We present a detailed study of pion and rho mass, decay constants
and quark mass
in Lattice QCD with two flavors of dynamical quarks. We use Wilson gauge and
fermion action at $\beta=5.6$ on $ 16^3 \times 32 $ lattice 
at eight values
of the Wilson hopping parameter in the range 0.156 - 0.158. 
We perform a detailed investigation of  the effect of
gaussian smearing on both source and sink. We determine the optimum
smearing parameter for various correlators for each
value of the Wilson hopping parameter. The effects of smearing on
observables are compared with those measured using local
operators. We also investigate systematic effects in the extraction of
masses and decay constants using different types of correlation
functions for pion observables. We make interesting observations
regarding chiral extrapolations and finite volume effects of different
operators.
 
\end{abstract}


\maketitle


\section{Introduction}
Among the various formulations of Lattice QCD, the Wilson 
formulation \cite{wilson} (both gauge and fermion action) proposed long ago is 
still attractive due to many reasons, first and foremost being the conceptual
simplicity. The Wilson formulation preserves 
discrete symmetries of the continuum formulation which simplifies the 
construction of 
lattice operators that correspond to the observables in the continuum theory. 
However, because of the explicit violation  of chiral symmetry by a dimension
five kinetic operator, Wilson formulation has been difficult to simulate at
light quark masses. Absence of chiral symmetry implies that the $``$physical'' 
quark mass is no longer proportional to the bare quark mass and the quark mass
renormalization is no longer only multiplicative.  Lack of chiral symmetry 
means that the
Wilson-Dirac operator which is a sum of anti-Hermitian and Hermitian terms is
not protected from arbitrarily  small eigenvalues and may lead to zero or near
zero modes for individual configurations. This is the infamous problem of 
$``$exceptional configurations''. This leads to convergence difficulties for
fermion matrix inversion which is an integral part of both the 
generation of gauge
field configurations with dynamical fermions and for the computation of 
hadronic
correlation functions via quark propagators. This poses difficulties for
lattice simulations with Wilson fermions in the chiral region. 
The situation has improved recently partly due to the
finding \cite{deldebbiojhep} that the numerical simulations are safe 
from accidental zero modes for large volumes. 

Very generally, the major challenges of probing the chiral regime of 
lattice QCD are: a) to be able
to achieve small lattice spacing $a$ to reduce scaling violations, b)
to have small current quark masses to have reliable chiral
extrapolations, and c) to have large enough physical volume of the
lattice to avoid finite size effects. In addition, lattice QCD
investigations suffer from uncertainties regarding the determination
of the lattice scale and inaccuracies in derived quantities of interest, e.g.,
hadronic masses, decay constants etc. Many of the above issues are
connected.

We have planned on a detailed lattice QCD investigation with Wilson
and Wilson-type fermions.
In this work we have taken the first step towards addressing the above
issues systematically. We study smearing of mesonic operators with
standard (unimproved) Wilson gauge and fermion actions with 2 fully dynamical
light quark flavors on $16^{3}32$ lattice, $\beta=6/g^2=5.6$ at 8 values
of bare quark masses (corresponding to fermionic hopping parameter
$\kappa =0.156,~ 0.1565,~0.15675, ~0.157, ~0.15725,
~0.1575,~0.15775,~0.158$). The lattice scale reached is respectable and
$a\sim 0.08$ fm ($a^{-1}\sim 2.45GeV$). It is important to have a
small enough $a$ to hopefully have small scaling violations and then
evaluate the low lying spectrum of QCD 
accurately and to study cleanly chiral extrapolations in finite
volumes. We plan
on having larger volumes and larger $\beta$ in the near future with
more sophisticated algorithmic developments discussed in
\cite{deldebbiojhep, deldebbio1, deldebbio2, japo}. In the
present work our emphasis is to have fully dynamical simulations at an
extensive set of $\kappa$ values and study accurate determinations of
the masses and decay constants in as many ways as possible.

In a separate paper \cite{qcd_paper1}, we investigate  quite elaborately the 
determination of
the lattice scale from the potential between a heavy quark-antiquark
pair. We study the observed change of $r_0/a$ with bare quark masses 
where $r_0$ is the Sommer parameter \cite{sommer} and resulting 
uncertainties of scale
determinations and also discuss its relation with the
chiral extrapolation of observables.  For a recent review on various
approaches to scale determination and associated issues, see 
Ref. \cite{mcneile}.
   
The masses in Lattice Gauge Theory are calculated from the 
asymptotic behaviour of Euclidean time correlation functions. The
contribution from the lowest mass state dominates for large time. Thus, to 
get a clean signal one has to make the length in the time direction as  
large as possible. But this is computationally expensive. Furthermore, as 
the time increases, the signal to noise ratio gets smaller which results in 
larger 
statistical errors. To reduce the statistical error, we need to increase 
the number of measurements (configurations). This is also computationally
expensive. Hence there is a practical need for techniques that allow one to 
reliably extract observables still working on moderately sized time direction 
and not too large number of configurations.    

For the measurement to be in the scaling region, one needs to go to smaller
and smaller lattice spacings. But as the lattice spacing decreases, the
physical hadron state will extend over more and more lattice spacings. Thus 
a state created by a local operator from the vacuum will have less and less
overlap with the physical state as the lattice spacing decreases.    
To improve the measurements, it is useful to use operators that have larger 
overlaps with the physical state (which is extended). This will naturally
lessen the contamination from higher mass states to the correlation
functions. Smearing of operators is one way to achieve this goal, thereby 
reducing the necessity to have both large time direction and large number of
configurations.  

One of the earliest references discussing the need for smearing is 
Ref. \cite{parisi}. Earliest suggestions for smearing involved  Dirac delta 
functions on source time slices, the so-called wall source. See for example,  
Ref. \cite{billoire}.
So far, in the literature, smearing has been implemented in two different 
methods, namely, the gauge invariant method and the  gauge fixed method. 
Gauge fixed wall source smearing, exponential smearing and gaussian smearing 
belong to the second method.
Gauge fixed wall source smearing is used in some work of 
JLQCD \cite{aoki-stagg} and MILC \cite{milc3} Collaborations.  
Exponential smearing is used in some works of CP-PACS \cite{alikhan} 
and JLQCD \cite{aoki} Collaborations.       
Gaussian smearing was first considered by 
DeGrand and Loft \cite{delo1}, 
and  used, e.g., in Ref.
\cite{bitar,hauswirth}) (with dynamical quarks) and in Ref.
\cite{ibm,milcw,degrand1,degrand2} (with quenched quarks) in different
contexts.
A clear discussion of smearing 
(in the context of gaussian smearing) and
the associated Fourier Transformation $``$trick" 
is provided in  the Ph. D. thesis of Hauswirth \cite{hauswirth}.

To the best of our knowledge, a systematic study of the effects of 
gaussian smearing on source or/and sink operators does not seem to be 
available in the literature.  
In this work we carry out a detailed investigation of the effect of gaussian
smearing on the source or/and sink meson operators. 

At a particular value of the smearing size,  
a variety of correlation functions are to be measured to 
extract meson observables. For example, for the pion, PP, AA, AP and PA
correlators can be used since both P and A carry quantum numbers of pion.
(Here P and A denotes psudoscalar and the fourth component of the 
axial vector densities respectively.) Mass gaps and amplitudes (coefficients)
are extracted for the smearing radius $s_0$ ranging from 1 to 8
in steps of 1. We also carry out extensive calculations with local operators. 
This helps us to quantify
the systematic effects in the determination of masses and decay constants.

We present results for the pion mass, the rho mass, their decay
constants and the quark mass (all in units of the lattice constant
$a$). Although this paper is not intended to deal with the issues of
chiral extrapolation of pseudoscalar observables, especially at our
relatively large masses with $16^3 32$ lattices, we make interesting 
observations regarding consistency with $\chi PT$. In this connection,
we also comment on finite volume effects on part of our data at the
largest $\kappa$. We have also noticed varying finite size effects on
different pion operators.
  
This paper is organized as follows. In Sec. \ref{simu} we present the
lattice action and the details of the simulation. Expressions for the
local observables that we calculate are given in Sec. \ref{obs_local}.
Sec. \ref{gauss_smear} introduces the gaussian smearing and different
fitting ansaetze for the correlator data analysis are given in
Sec. \ref{analysis}. Details of the implementation of smearing are in
Sec. \ref{smearing_details}. Results for pion and rho observables are 
presented in Secs. \ref{results_pi} and  \ref{results_rho}
respectively. Finally, the summary and the conclusions are presented
in Sec. \ref{conclu}. For the sake of completeness and clarity, the
Fourier transform method employed in the case of sink smearing is
presented in Appendix \ref{ft_trick}.      

\section{Simulation}\label{simu}
\subsection{Action}
We have performed simulations with the standard Wilson action
$S ~=~ S_{F} ~+~ S_{G}$ where the standard Wilson gauge action is given by
$$ S_{G} =
~ \beta~ \sum \left [1-\frac{1}{3}~{\rm Re}~{\rm Trace}
~U_P \right ], $$
with $ \beta = \frac{6}{g^2}$ ($g$ is the $SU(3)$ gauge coupling), the
elementary 
plaquette $U_P$ being the  product of $SU(3)$
link fields $U$ around the elementary square of the hypercubic lattice
and the standard Wilson fermion action is given by
\begin{eqnarray}
 S_F[\psi,{\overline\psi},U] =
  \sum_{x,y}{\overline \psi}_{x} M_{xy}\psi_{y}~~ {\rm with}~~
\end{eqnarray}

\begin{eqnarray}
M_{xy} = \delta_{xy} -
  \kappa\left[\left(r-\gamma_\mu\right)U_{x,\mu }~\delta_{x+\mu,y}  +
\left(r + \gamma_\mu\right)U^\dagger_{x-\mu, \mu}~\delta_{x-\mu,y} \right]~.
\end{eqnarray}
We have suppressed spin, color and flavor indices in the quark fields
$\psi$ and $\overline \psi$ and the fermion matrix $M$. As usual, 
we have taken the Wilson parameter $r=1$. 
We consider two degenerate light quark flavors, i.e., $N_F=2$.

\subsection{Details of Simulation}
The gauge coupling $\beta=5.6$ and the lattice volume is 
$ 16^3 \times 32$. The hopping parameter $\kappa$=0.156, 0.1565, 0.15675, 
0.157, 0.15725, 0.1575, 0.15775 and 0.158. At each $\kappa$
we have generated 5000 equilibrated configurations 
with the standard HMC algorithm (with even-odd pre-conditioned
Conjugate Gradient for inversion of $M^{\dagger}M$)
and performed the mesonic correlator measurements separated by 
25 trajectories. The time step size is chosen to be 0.01 and the number 
of steps per trajectory is 100. The stopping criterion for the residue
of the conjugate gradient  
$ \sqrt{\mid R\mid^2/\mid \chi\mid^2}\leq 10^{-5}$
 is used for all $\kappa$ values except 0.158 (for which 
it is $10^{-6}$) where $ R = M^{\dagger}M\zeta-\chi$ ($\chi$ and
$\zeta$ are respectively the source and the iterative solution).
Stabilized BICG algorithm is used for the inversion of $M$ needed for
the mesonic correlators, 
the stopping criterion of the residue for very accurate inversion in
this case is chosen as $10^{-9}$ and $10^{-10}$.

The parameters are chosen such that the acceptance rate is above 75 \% 
for $\kappa$ less than 0.157 and around 65 \% for $\kappa$ above and 
including 0.157.  

\vskip .1in
\begin{figure}
\begin{center}
\includegraphics[width=.6\textwidth]{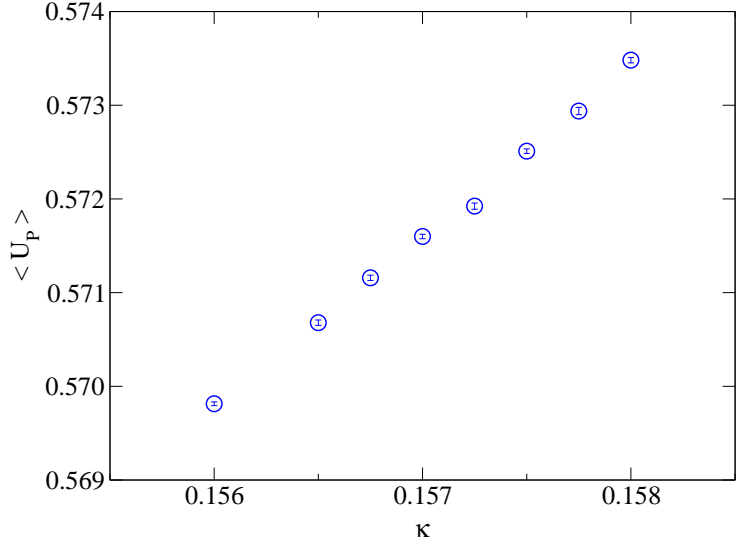}
\end{center}
\caption{Average value of the Plaquette versus $\kappa$. }
\label{pvsk}
\end{figure}

Fig. \ref{pvsk} shows that the expectation value of the average
plaquette has a smooth behavior in dependence of $\kappa$.
    
\subsection{Auto correlation}
For any particular observable ${\mathcal{O}}$, autocorrelations among the
generated configurations are generally determined by the integrated
autocorrelation time $\tau_{\rm int}^{\mathcal{O}}$ for that
observable. For this purpose, at first, one needs to calculate the
unnormalized autocorrelation function of the observable ${\mathcal{O}}$
measured on a sequence of $N$ equilibrated configurations as  
\begin{equation}
C^{\mathcal{O}}\left(t\right) = \frac{1}{N - t}\sum_{r = 1}^{N - t}
\left({\mathcal{O}}_r - \langle{\mathcal{O}}\rangle_L\right)
\left({\mathcal{O}}_{r+t} - \langle{\mathcal{O}}\rangle_R\right)
\end{equation}
where, instead of the usual single mean-value estimator, we have used the
``left'' and ``right'' mean-value estimators
\begin{equation}
\langle{\mathcal{O}}\rangle_L = \frac{1}{N - t}\sum_{s = 1}^{N - t}
{\mathcal{O}}_s,~~~~ 
\langle{\mathcal{O}}\rangle_R = \frac{1}{N - t}\sum_{s = 1}^{N - t}
{\mathcal{O}}_{s+t}
\end{equation} 
for a faster convergence as claimed in \cite{orthprd}.

 Following the ``windowing'' method as recommended by Ref.
 \cite{sokal}, we have calculated the integrated autocorrelation time as   
\begin{equation}
\tau_{\rm int}^{\mathcal{O}} = \frac{1}{2} + \sum_{t=1}^{t_{\rm cut}}
\rho^{\mathcal{O}}\left(t\right)
\label{taudef}
\end{equation}
where 
\begin{equation}
\rho^{\mathcal{O}}\left(t\right) = C^{\mathcal{O}}\left(t\right)/
C^{\mathcal{O}}\left(0\right)
\end{equation}
is the normalized autocorrelation function.

The factor of $\frac{1}{2}$ in the Eq. (\ref{taudef}) is purely a matter
of convention and is chosen such as to guarantee that for
$N\rightarrow\infty$, the normalized autocorrelation function behaves as
${\rm exp}\left(-t/\tau_{\rm int}^{\mathcal{O}}\right)$ with
$\tau_{\rm int}^{\mathcal{O}}\gg1$. 

If plotted against the variable cutoff $t_{\rm cut}$, the resulting values of
$\tau_{\rm int}^{\mathcal{O}}$ ideally show a plateau but instead a peak or
monotonous rise is also observed sometimes. Following the suggestions
of \cite{sokal}, we have drawn two straight lines $t_{\rm cut}/4$ and 
$t_{\rm cut}/10$ respectively and have chosen $\tau_{\rm
  int}^{\mathcal{O}}$ inspecting only the segment of the curve lying within its
intersections with the two straight lines. This ensures a balance
between the noise and the bias in the estimation of the autocorrelation time.

We have measured the integrated autocorrelation times for the average values of 
the Plaquette and the number of Conjugate Gradient iterations, called   
${\rm Niter}$,  during HMC trajectories for the generated 
configurations for each value of $\kappa$. The errors are calculated
by the single omission jackknife method. 
Our results are presented in the Table \ref{auto}. Despite
fluctuations, $\tau_{\rm int}$ in general increases with larger  
$\kappa$ especially as $\kappa$ approaches $\kappa_c$.   

\begin{figure}
\begin{center}
\includegraphics[width=.8\textwidth]{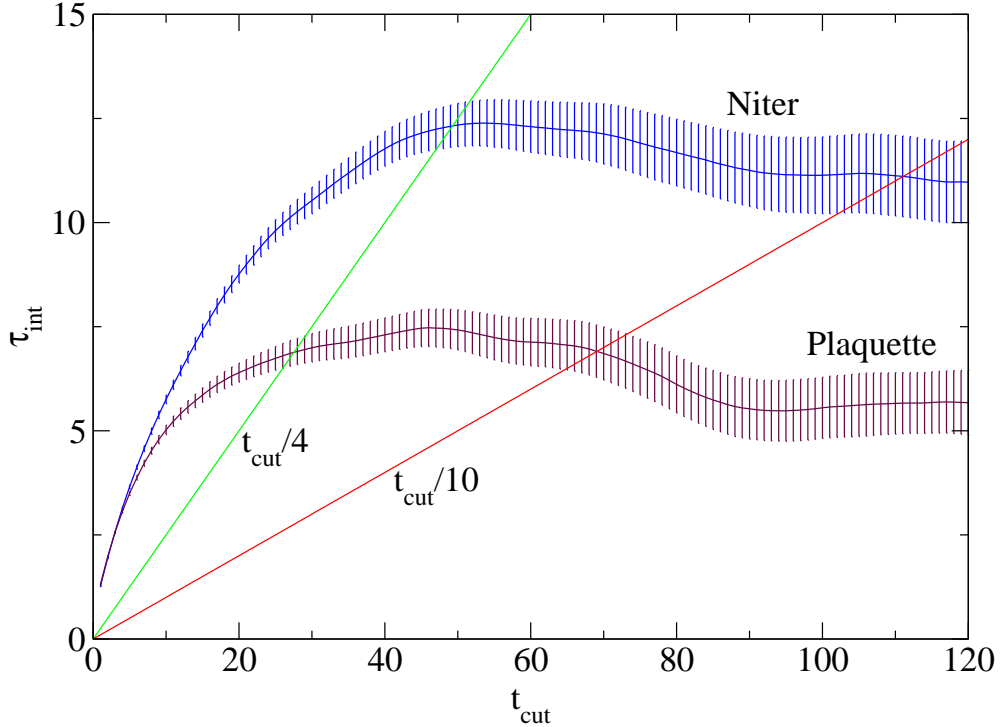}
\end{center}
\caption{ Integrated autocorrelation time $\tau_{\rm int}$ for both average 
Niter and Plaquette at $\kappa$=0.157. }
\label{autocorr}
\end{figure}

A typical example of the expected plateau of $\tau_{\rm int}$ as a
function of $t_{\rm cut}$ for both the average Niter and Plaquette at
$\kappa=0.157$ is shown in Fig. \ref{autocorr}. 

\vskip .01in
\vskip .1in
\begin{table}
\begin{tabular}{|l|l|l|}
\hline \hline
$\kappa$ & {Niter} & {Plaquette}\\
\hline
{0.156}&{13(1)} &{5.4(3)} \\
\hline 
{0.1565}&{14(1)} &{11(1)} \\
\hline
{0.15675}&{19(1)} &{8(1)} \\
\hline
{0.157}&{12(1)} &{7(1)} \\
\hline
{0.15725}&{26(1)} &{11(1)} \\
\hline
{0.1575}&{21(1)} &{8(1)} \\
\hline
{0.15775}&{14(1)} &{9(1)} \\
\hline
{0.158}&{27(2)} &{10(1)} \\
\hline\hline
\end{tabular}
\caption{Integrated Autocorrelation time $\tau_{int}$ for Average Niter and 
Plaquatte as a function of $\kappa$.}
\label{auto}
\end{table}

\section{Observables with local operators}\label{obs_local}
We measure the charged pion and the rho propagators to extract masses, 
decay constants and the quark mass at each $\kappa$.

For pion, we measure the following zero-spatial-momentun correlation 
functions on a $L^3T$ lattice as functions of the Euclidean time $t$:
\begin{eqnarray*}
& & C_1(t)~~ =~~\langle 0 \mid {\cal O}^\dagger(t) {\cal O}(0) 
\mid 0 \rangle ~~ \stackrel{t \rightarrow \infty}{\longrightarrow}~~ 
C^{\cal OO} \Big [ e^{-m_\pi t}~ +~ e^{-m_\pi (T-t)} \Big ]   \\
& & C_2(t)~~ =~~\langle 0 \mid {\cal O}_1^\dagger(t) {\cal O}_2(0) \mid 
0 \rangle ~~ \stackrel{t \rightarrow \infty}{\longrightarrow}~~ 
C^{{\cal O}_1{\cal O}_2} \Big [ e^{-m_\pi t}~ -~ e^{-m_\pi (T-t)} \Big ]  \\
\end{eqnarray*}
The coefficients are given by,
\begin{eqnarray*}
& & C^{\cal OO} = \frac{1}{2 m_\pi} \mid \langle 0 \mid {\cal O}(0) 
\mid \pi \rangle
\mid^2 \\
& & C^{{\cal O}_1{\cal O}_2}~ = ~\frac{1}{2 m_\pi} \langle 0 
\mid {\cal O}_1^\dagger(0) \mid \pi \rangle 
 \langle \pi \mid {\cal O}_2(0) \mid 0 \rangle~.
\end{eqnarray*}

${\cal OO}\equiv PP$ or $AA$ and ${{\cal O}_1{\cal O}_2}= AP$ or $PA$ 
where $P = {\overline q}_i \gamma_5 q_j~~~ {\rm and}~~ 
A_4 = {\overline q}_i \gamma_4 \gamma_5 q_j~$  denote the pseudoscalar
density and fourth component of the axial vector current ($i$ and $j$ stand for
flavor indices for the $u$ and $d$ quarks, for the charged pion $i\ne
j$).

For clarity and to set up our notation, in the following we discuss
{\em all} the possible ways of determining the decay constants and
the PCAC quark masses.


\subsection{Pion Decay Constant}

The pion decay constant $F_\pi$ and the quark mass $m_q$ from PCAC or 
the axial Ward identity are  respectively defined, in the
continuum, via 
\begin{eqnarray}
\langle 0\mid A_\mu(0)\mid \pi(p)\rangle &=& \sqrt{2} F_\pi p_\mu, \label{fpi}\\
\partial_\mu A_\mu (x) &=& 2 m_q P(x). \label{PCAC}
\end {eqnarray}

Since we measure the PP, PA, AP and AA correlators, we have a variety of ways 
to compute the pion decay constant and the PCAC quark mass. 

\vskip .1in
{\bf Method I}: From the AA correlator  
\begin{eqnarray}
C(t) ~&=&~\langle 0 \mid A_4^\dagger(t) A_4(0) \mid 0 \rangle ~\\
&\stackrel{t \rightarrow \infty}{\longrightarrow}& 
C^{AA}\Big [ e^{-m_\pi t}~ +~ e^{-m_\pi
  (T-t)} \Big ]  
\end{eqnarray}
where
$C^{AA} = \frac{1}{2 m_\pi} \mid \langle 0 \mid A_4(0) \mid \pi \rangle
\mid^2$,
and the pion decay constant as defined in Eq. (\ref{fpi}) follows 
\begin{eqnarray}
 F_{\pi}^{AA} ~ = ~ 2\kappa ~ \sqrt{\frac{C^{AA}}{m_\pi}}~. \label{fpiAA}
\end{eqnarray}
The factor $2\kappa$ above accounts for the difference in normalization 
between the continuum and the Wilson lattice fermion actions.

It is obvious that the $m_{\pi}$ that appears in Eq. (\ref{fpiAA}) is
numerically
obtained from the $AA$ correlator. If numerically $m_{\pi}$ is obtained from
another correlator, say the PP, then one needs to put that mass in the fit
Ansatz for the AA correlator to obtain the coefficient $C^{AA}$. 

\vskip .1in
{\bf Method II}: 
From the PP and the AP propagators
\begin{eqnarray}
C^{PP} &=&  \frac{1}{2 m_\pi} \mid \langle 0 \mid P(0) \mid \pi \rangle \mid^2 \\
C^{AP} &=& \frac{1}{2 m_\pi}  \langle 0 \mid A_4(0) \mid \pi \rangle
\langle \pi \mid P^{\dagger}(0) \mid 0 \rangle
\end{eqnarray}
which lead to
\begin{eqnarray}
 F_\pi^{AP} ~ = ~ \frac{2\kappa ~ C^{AP}}{\sqrt{ m_\pi C^{PP}}} \label{fpiAP}.
 \end{eqnarray}
 Similarly using the PP and the PA propagators
 \begin{eqnarray} 
 F_\pi^{PA} ~ = ~ \frac{2\kappa ~ C^{PA}}{\sqrt{ m_\pi C^{PP}}}. \label{fpiPA}
\end{eqnarray}

For numerical evaluation of $F_{\pi}$ using Eqs. (\ref{fpiAP}) and
(\ref{fpiPA}), the values of $m_{\pi}$ computed from PP, AP and PA
correlators have to be the same, which is hard to achieve numerically. Hence 
it is advisable to use the best determined mass from one particular correlator 
and then use that value in the other correlators 
to determine all the coefficients $C$. The same applies also to the 
determination of the different PCAC quark masses discussed below.

\subsection{ PCAC Quark Mass}

{\bf Method I}: From $ \partial_\mu A_\mu(x) = 2 m_q P(x), $ summing over
spatial coordinates,
\begin{eqnarray}
\sum_{\bf x} \partial_\mu A_\mu(x) = 
2 m_q \sum_{\bf x} P(x) \Longrightarrow 
\partial_4 A_4(t) = 2 m_q P(t).
\end{eqnarray}
Taking the matrix element between the vaccum and physical pion states we have
\begin{eqnarray}
m_\pi \langle 0 \mid A_4(0)\mid \pi \rangle = 
2 m_q \langle 0 \mid P(0)\mid \pi \rangle 
\end{eqnarray}
which leads to 
\begin{eqnarray}
& & m_q^{AA} ~ = ~\frac{m_\pi}{2} ~ \sqrt{\frac{C^{AA}}{C^{PP}}}~. 
\end{eqnarray}
\vskip .1in
{\bf Method II}: Using PCAC
\begin{eqnarray}
\partial_\mu \langle 0 \mid A_\mu (x)P^\dagger(0) \mid 0 \rangle =
2 m_q \langle 0 \mid P (x)P^\dagger(0) \mid 0 \rangle
\end{eqnarray}
Summing over spatial coordinates 
\begin{eqnarray}
\sum_{\bf x} \partial_\mu \langle 0 \mid 
A_\mu (x)P^\dagger(0) \mid 0 \rangle =
2 m_q \sum_{\bf x} 
\langle 0 \mid P (x)P^\dagger(0) \mid 0 \rangle
\end{eqnarray}
At large $t$,
\begin{eqnarray}
\partial_4 C^{AP}\Big [ e^{-m_\pi t} - e^{-m_\pi (T-t)}\Big ] = 2 m_q C^{PP}
\Big [ e^{-m_\pi t} + e^{-m_\pi (T-t)}\Big ]
\end{eqnarray}
which leads to 
\begin{eqnarray}
 m_q^{AP} = \frac{m_\pi}{2} ~ \frac{C^{AP}}{C^{PP}}, ~~~~~~~~~~~~ 
m_q^{PA} = \frac{m_\pi}{2} ~ \frac{C^{PA}}{C^{PP}}
\end{eqnarray}

\subsection{Rho Mass and  Decay Constant}

There are two different definitions of $\rho$ deacy constant used in 
the literature.
\begin{eqnarray}
\langle 0 \mid V_\mu(0) \mid \rho \rangle ~ = ~ \epsilon_\mu ~ 
\frac{m_\rho^2}{f_\rho}
\end{eqnarray}
where $f_\rho$ is dimensionless and 
\begin{eqnarray}
\langle 0 \mid V_\mu(0) \mid \rho \rangle ~ = ~ 
\epsilon_\mu \sqrt{2}F_\rho m_\rho
\end{eqnarray}
with $ V_\mu(0) = {\overline q}(0) \gamma_\mu q(0) $ and $\epsilon_\mu$ 
is the polarization vector of rho.

Here $ F_\rho$ is dimensionful and $ F_\rho/m_\rho = (f_\rho)^{-1} $.

Mass and decay constant of rho are calculated from the correlation function
\begin{eqnarray}
C(t) ~&=&~\langle 0 \mid V_3^\dagger(t) V_3(0) \mid 0 \rangle ~\\
&\stackrel{t \rightarrow \infty}{\longrightarrow}& 
C^{VV}\Big [ e^{-m_\rho t}~ +~ e^{-m_\rho
  (T-t)} \Big ]  
\end{eqnarray}
where,
$C^{VV} = \frac{1}{2 m_\rho} \mid \langle 0 \mid V_3(0) \mid \rho \rangle
\mid^2$. Thus
\begin{eqnarray}
F_\rho/m_\rho =  1/f_\rho ~ = ~ 2\kappa ~ \sqrt{\frac{ C^{VV}}{m_\rho^3}}~.
\end{eqnarray}

{\em We compute the decay constants and the quark masses according to the
expressions given above, but to obtain their values in the continuum
one needs to multiply with appropriate factors of renormalization
constants $Z_P$, $Z_A$ and $Z_V$ associated with lattice pseudoscalar,
axial vector and vector densities respectively}. Although we have made 
approximate estimations of $Z_A$ and $Z_V$ in Sec. \ref{results_rho}, we do
{\em not} quote numbers in the continuum in this paper.

\section{Gaussian Smearing} \label{gauss_smear}

To increase the overlap with the hadronic ground state, many, if not most,
QCD spectrum calculations traditionally use the method of smearing the
hadronic interpolating operator, essentially making the hadronic operator
spread around their central location in space. In this work,   
for the pion and the rho operators, we have used the so-called gaussian smearing
where one uses a shell model trial
wave function with one variational parameter, $\phi(r) \sim \exp
\left(-\left(r/s_0\right)^2\right)$. For details please see Ref.
[\cite{delo1}]. $s_0$ is the smearing size parameter.

An advantage of the gaussian smearing is that the smearing function, for
example for a meson operator, separates into two factors one belonging to
the quark and the other to the antiquark. This has certain numerical
advantages, especially for sink smearing. However, the smeared operators are
no longer gauge-invariant because the quark and the antiquark are spatially
separated. We have used Coulomb gauge fixing to obtain non-zero expectation
values.   

The coefficients $C^{{\cal O}_1{\cal O}_2}$ for local operators 
are needed, as described in
Sec. \ref{obs_local} to calculate the decay constants and the PCAC 
quark masses. 
One can take a mixed approach where one evaluates the hadron mass from a
hadronic propagator which uses smearing at either the source or the
sink or at both places, and at the same time calculate the coefficient
$C^{{\cal O}_1{\cal O}_2}$ from the local-local propagator (as pursued in
Ref. \cite{milc3}). For this to work, large time-extents are
necessary. However, if one wants to calculate the local-local coefficients
from the smeared propagators, 
one needs to calculate the propagators with {\em all} combinations 
of smearing: i)
local sink and smeared source ($ls$), ii) smeared sink and local source
($sl$), and (iii) smeared sink and smeared source ($ss$). If they produce
the same hadronic mass at large euclidean times, by combining the
coeffecients of the three, all calculated with the same smearing parameter
$s_0$, one is able to calculate the coefficient
corresponding to the local-local propagator. This is done as follows.  
 
The large Euclidean time behavior of the correlation function
involving local (unsmeared) operators is 
${\cal A}$ and ${\cal B}$
\begin{eqnarray*}
\langle 0 \mid {{\cal A}_l^{\rm sink}}^\dagger (t)~ 
{{\cal B}_l^{\rm source}} (0) 
\mid 0 \rangle 
~ & \stackrel{t \rightarrow \infty}{\longrightarrow} &  
C^{\cal AB}_{ll}~  e^{-m_Ht} 
\end{eqnarray*} 
where, 
$C^{\cal AB}_{ll} ~ = ~ ~\frac{1}{2 m_H}
\langle 0 \mid {{\cal A}_l^{\rm sink}}^\dagger (0)\mid H \rangle \langle H
\mid~ {{\cal B}_l^{\rm source}} (0) \mid 0 \rangle~$.

Similarly
\begin{eqnarray*}
C^{\cal AB}_{ls} ~ &=& ~ ~\frac{1}{2 m_H}
\langle 0 \mid {{\cal A}_l^{\rm sink}}^\dagger (0)\mid H \rangle \langle H
\mid~ {{\cal B}_s^{\rm source}} (0) \mid 0 \rangle~ \nonumber \\
C^{\cal AB}_{sl} ~ &=& ~ ~\frac{1}{2 m_H}
\langle 0 \mid {{\cal A}_s^{\rm sink}}^\dagger (0)\mid H \rangle \langle H
\mid~ {{\cal B}_l^{\rm source}} (0) \mid 0 \rangle~ \nonumber \\
C^{\cal AB}_{ss} ~ &=& ~ ~\frac{1}{2 m_H}
\langle 0 \mid {{\cal A}_s^{\rm sink}}^\dagger (0)\mid H \rangle \langle H
\mid~ {{\cal B}_s^{\rm source}} (0) \mid 0 \rangle~ .
\end{eqnarray*}

Assuming the lowest mass obtained at large Euclidean time to be the same
for each of the $ls$, $sl$ and $ss$ correlators, it follows that
\begin{equation}
C_{ll} = C_{ls}C_{sl}/C_{ss}~. \label{ls_sl_by_ss} 
\end{equation}

\section{Analysis of Meson Correlation Functions} \label{analysis}
For our initial investigations we performed four types of fits using 
exponential functions for the correlation functions of the pion and the rho. 
For example, for the pion propagator, the fit ansaetze involving two, three,
four and five parameters are as follows. 

\begin{eqnarray}
C_{I}(t) &=& C_1^{{\cal O}_1{\cal O}_2} \Big [ e^{-m_{\pi}t} ~ \pm ~ e^{-m_{\pi}(T- t)} \Big ]~, \nonumber\\
C_{II}(t) &=& C_1^{{\cal O}_1{\cal O}_2} \Big [ e^{-m_{\pi}t} ~ \pm ~ e^{-m_{\pi}(T- t)} \Big ] ~+~
C_2^{{\cal O}_1{\cal O}_2} \Big [ e^{-3m_{\pi}t} ~ \pm ~
  e^{-3m_{\pi}(T- t)} \Big ]~, \nonumber \\
C_{III}(t) &=& C_1^{{\cal O}_1{\cal O}_2} \Big [ e^{-m_{\pi}t} ~ \pm ~ e^{-m_{\pi}(T- t)} \Big ] ~+~
C_2^{{\cal O}_1{\cal O}_2} \Big [ e^{-m_2t} ~ \pm ~ e^{-m_2(T- t)}
  \Big ]~, \nonumber \\
C_{IV}(t) &=& C_1^{{\cal O}_1{\cal O}_2} \Big [ e^{-m_{\pi}t} ~ \pm ~ e^{-m_{\pi}(T- t)} \Big ] ~+~
C_2^{{\cal O}_1{\cal O}_2} \Big [ e^{-3m_{\pi}t} ~ \pm ~ e^{-3m_{\pi}(T- t)} \Big ] \nonumber \\
 &+&C_3^{{\cal O}_1{\cal O}_2} 
\Big [ e^{-m_3t} ~ \pm ~ e^{-m_3(T- t)} \Big ]~. \label{expfit}
\end{eqnarray} 

In the presence of sea quarks, pair creation from vacuum becomes possible 
and the creation of two quark-antiquark pairs results in two extra 
pseudoscalar mesons. Thus in the presence of sea quarks,  
the next higher state can be expected to be a three pion state 
\cite{deldebbio1}. 
The second and the fourth fit ansaetze above
are motivated by this physical picture. The rho propagator can also
be treated in the same way, the next higher state there being
$2m_{\pi}+m_{\rho}$.  

\begin{figure}
\centering
\includegraphics[width=4in,clip]{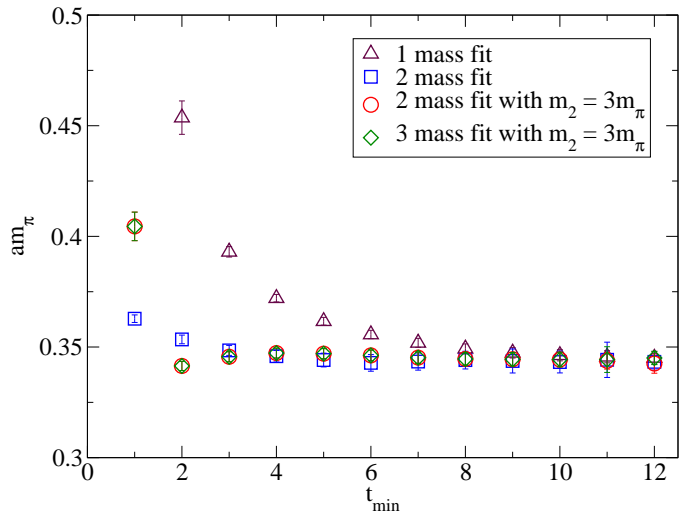}
\caption{Pion mass obtained from unsmeared PP correlator at $\kappa = 0.157$   
with four different fits as function of $t_{\rm min}$.}
\label{mpiPPll_vs_tmin}
\end{figure}



\section{Details of the implementation of the gaussian smearing} 
\label{smearing_details}


\begin{figure}
\centering
\includegraphics[width=4in,clip]{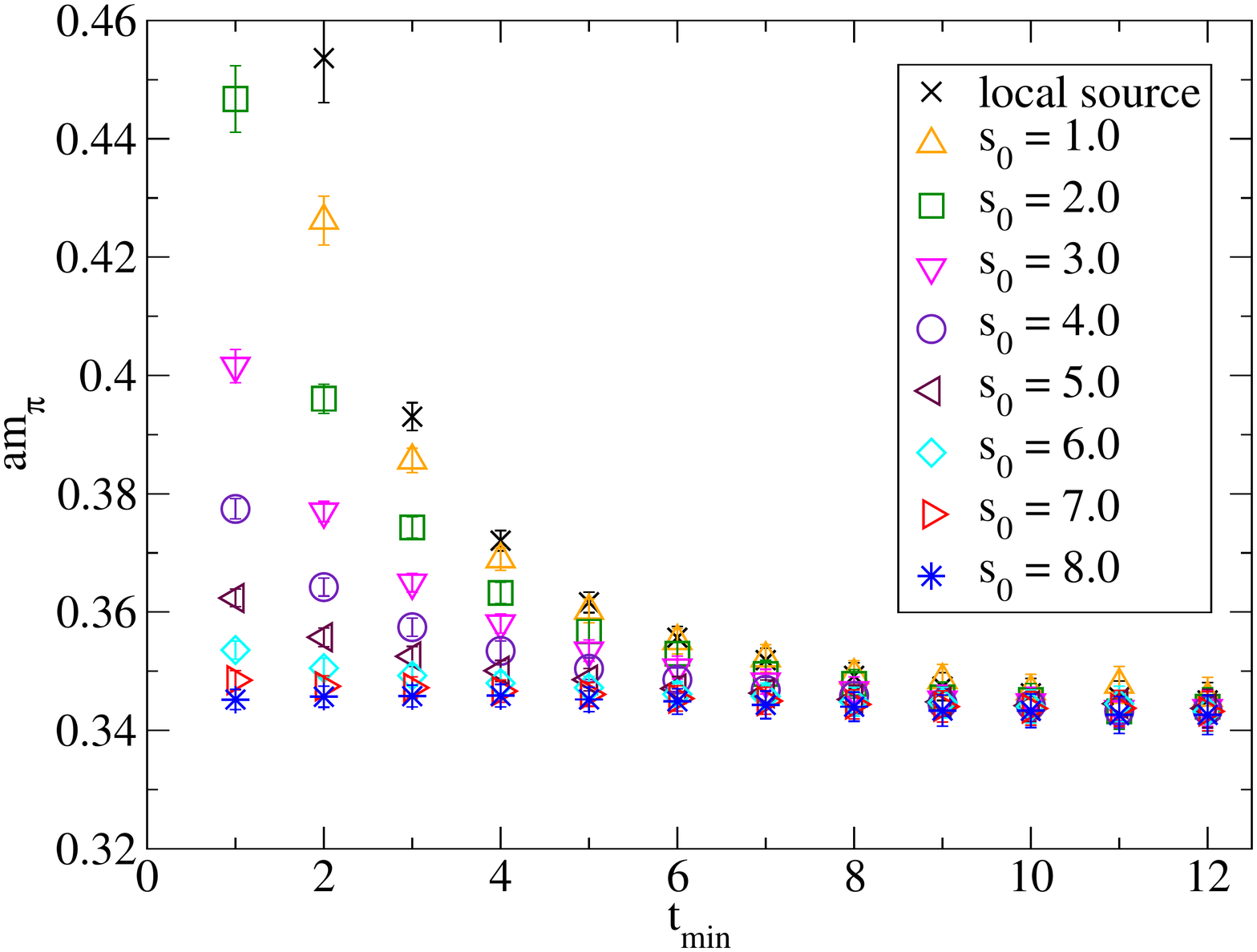}
\caption{Pion mass from one exponential fit (two parameters) as a 
function of $s_0$.}
\label{mpr0}
\end{figure}

\begin{figure}[tp]
\centering
\includegraphics[width=4in,clip]{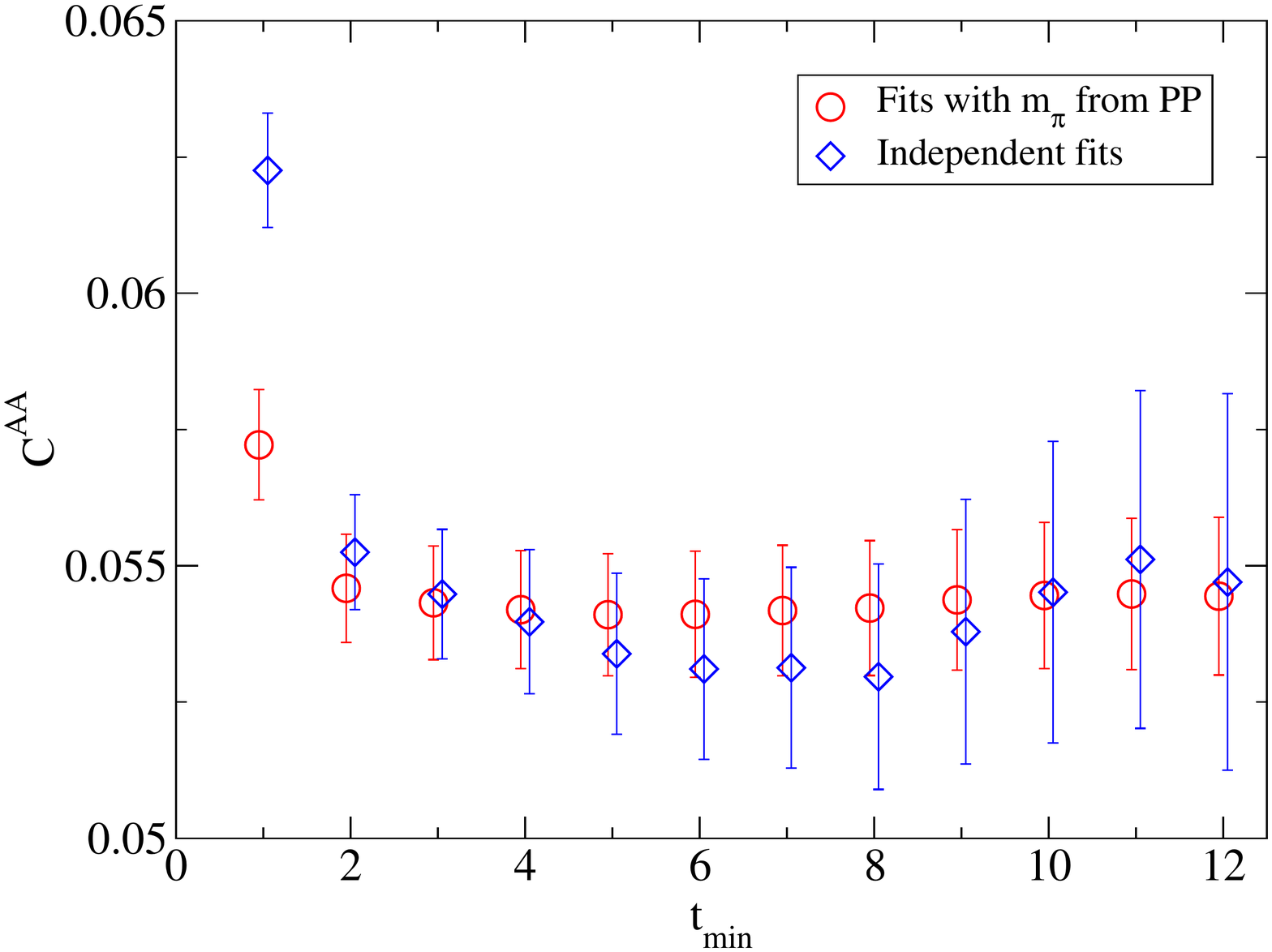}
\caption{$C_{AA}$ obtained from constrained and unconstrained
  fits (data slightly shifted horizontally for clarity on the error bars).}
\label{coefcomp}
\end{figure}

As mentioned earlier, in this paper we present results with two flavors of
fully dynamical Wilson (unimproved) quarks (with standard plaquette Wilson
gauge action) at $\beta=5.6$ on $16^3 32$ lattices at an extensive set of
the fermion hopping parameter, viz., $\kappa=0.156,~0.1565,~0.15675,~0.157,
~0.15725,~0.1575,~0.15775,~0.158$.

At this $\beta$ and lattice volume, similar $N_F=2$ calculations have been
done before at a few $\kappa$ values (for example, see Ref. \cite{orthprd}). 
The previous results help us to have a cross-check and have belief in our
numerical procedure. Some data are available 
at larger volumes with the rest of the parameters staying the same; these
give some indication of the finite size effects in our results.
However, we like to mention that, to the best of our
knowledge, ours are the first calculations at $\beta=5.6$ at $\kappa=0.15675,
~0.15725~~{\rm and}~~0.15775$ for any lattice volume. 

At each $\kappa$, we have investigated in detail the PP, AA, AP and PA 
correlators for the pion and VV correlators for the rho. For each operator,
we have used source smearing ($ls$), sink smearing ($sl$) and smearing at both
source and sink ($ss$) (see Appendix \ref{ft_trick} for the Fourier
transform method for
the smeared sources). Then at each $\kappa$, with each operator and each
sink-source smearing combination, we have tried eight values for the
gaussian smearing size parameter, viz., $s_0 = 1$ to $L/2=8$ 
(in increment of unity) in order to have
optimum smearing (please note that $s_0\rightarrow \infty$ produces the
so-called wall source). 

For comparison, we have also investigated all the above correlators without
any smearing, i.e., using local sink and local source (denoted with
subscript $ll$).

With or without smearing, our general strategy was to determine the pion
mass from the PP and the AA propagators only and not from the asymmetric
propagators AP or PA, by reaching a plateau with
respect to $t_{\rm min}$ (minimum $t$ used for fitting the
formulae \ref{expfit}. In general, the quality of the mass-plateaux for 
the AP and the PA correlator was not of the same high quality as for the PP 
and the AA correlators (AA noisier than PP), more true with smearing. As a
note of caution, let us also mention that at a few large $\kappa$ values
the local-local AA correlators were very noisy, 
resulting in inaccurate and perhaps
unreliable determinations of the all relevant quantities as apparent from
fourth data-column of Table \ref{piondata}.    

We have used the well-known technique of effective masses by taking ratios
of propagators for adjacent time-slices, but only as a rough guide (especially
for smeared propagators). We shall {\em not} show any effective mass plots
in this paper.

In Fig. \ref{mpiPPll_vs_tmin}, we show, for the unsmeared ($ll$) PP
correlator, the results at $\kappa=0.157$ for $m_{\pi}$ obtained with the four 
different fit Ansaetze \ref{expfit}, as functions of $t_{\rm min}$ (here
and in the subsequent figures, $t_{\rm min}/a$ in
the abscissae is indicated as just $t_{\rm min}$. Firstly, we
note in Fig. \ref{mpiPPll_vs_tmin} that all the different fits produce
good plateaux for appropriately large $t_{\rm min}$ and
they all agree with each other. The plot also shows that $3m_{\pi}$ is the
next higher state. Unless otherwise specified, all errors in this plot and 
others to follow are single-omission jackknife
statistical errors calculated from 200 jackknife bins of correlator data.

With unsmeared propagators, although it is obvious that longer time-extent
would have helped, we have been quite successful in obtaining a reasonable 
plateau even at lower pion masses with the single exponential ansatz.
This was a great advantage because ansaetze with higher number of 
exponentials are harder to automate for the
error calculation because sometimes the order of the masses changes in the 
fit-results for some bins. We have (almost) always used the single 
exponential fit for the unsmeared correlators both for the pion and the rho.    
              
Once we determine the pion mass from the PP and the AA correlators, we then 
respectively use these masses for fitting all the correlators to obtain the 
coefficients, as discussed in Sec. \ref{obs_local} and then extract the various
decay constants and the quark masses. We have used the same strategy for 
the smeared propagators too, both for the pion and the rho.

Fig. \ref{mpr0} shows for $\kappa=0.157$ the pion mass determined from 
single-exponential fits to the PP correlator with local sink and smeared source
as a function of $t_{\rm min}$ for 
all the investigated values of the smearing size parameter $s_0=1, 2,\ldots,8$.
The figure also includes the pion mass obtained by single-exponential fits 
from the local-local PP correlator for this $\kappa$. The figure shows 
that $s_0=8$ gives the most reliable mass-plateau and we take the mass from 
the fit with the best confidence level from the plateau at $s_0=8$. As 
discussed below, choice of the optimum $s_0$ for a given correlator, 
PP in this 
case, has to be the same for all smearing combinations $ls$, $sl$ and $ss$.
Hence in this case we made sure that for the other two cases, viz., $sl$ and 
$ss$, the optimum choice of $s_0$ was also 8. Then along with $C^{PP}_{ls}$ 
obtained from the correlated 2-parameter single-exponential fits to the 
$ls$ correlator, we also obtained
$C^{PP}_{sl}$ and $C^{PP}_{sl}$ at $s_0=8$ by making 1-parameter single 
exponential fit with the value of $m_{\pi}$ put in from the chosen best fit to
the $ls$  PP correlator. This process was done for each of 200 jackknife bins
so as to get the jackknife statistical errors for all the derived quantities.
The errors shown in Fig. \ref{mpr0} are jackknife errors.       

Once $m_{\pi}$ is determined from a given correlator (in this $ls$
PP), we can then similarly determine the coefficients
for all other correlators (in this case, AA, AP and PA) and for 
all sink-source smearing combinations. {\em Optimum values of the smearing
size $s_0$ need not be the same for the other operators}. However, it
has to be the same for all smearing combinations of the same operator
(for applicability of Eq. (\ref{ls_sl_by_ss}). Once 
again, for each correlator we looked for the optimum $s_0$ by comparing 
1-parameter single-exponential fits (with $m_{\pi}$ put in) for all smearing 
combinations. In Fig. \ref{coefcomp} the coefficient $C^{AA}_{ls}$ for the 
optimum $s_0=3$ for the AA correlator is shown as a function of
$t_{\rm min}$ 
again at $\kappa=0.157$ and displays a nice plateau with accurate data points. 
The figure also compares $C^{AA}_{ls}$ obtained from
correlated 2-parameter single-exponential fit (with no $m_{\pi}$ 
predetermined); coefficients determined in this 
way show a rough plateau with a lot less accuracy (larger error bars). This is 
an important point considering that the determination of quark masses and 
especially the decay constants depend crucially on the  
accurate determination of the coefficients.

\begin{table}
\begin{tabular}{|l|l|l|l|l|l|}
\hline \hline
$\kappa$ &\multicolumn{5}{c|}{$s_0$} \\
\cline{2-6}
&  {PP} & {AA} & {AP} & {PA}& {VV}\\
\hline
{0.156}&{3.0} &{1.0} &{1.0} &{1.0} &{4.0}\\
\hline 
{0.1565}&{7.0} &{5.0} &{4.0} &{5.0} &{7.0}\\
\hline
{0.15675}&{5.0} &{3.0} &{2.0} &{2.0} &{7.0}\\
\hline
{0.157}&{8.0} &{3.0} &{5.0} &{4.0} &{8.0}\\
\hline
{0.15725}&{6.0} &{3.0} &{3.0} &{3.0}  &{8.0}\\
\hline
{0.1575}&{6.0} &{3.0} &{3.0} &{3.0} &{8.0}\\
\hline
{0.15775}&{8.0} &{4.0} &{5.0} &{4.0} &{8.0}\\
\hline
{0.158}&{7.0} &{7.0} &{5.0} &{5.0} &{8.0}\\
\hline\hline
\end{tabular}
\caption{Optimum values of the smearing size parameter $s_0$ for all correlators investigeted}\label{r0}
\end{table}
\vskip 0.2in  

As mentioned already before while discussing the local-local correlators, the 
pion masses were always determined from only the PP and the AA correlators
also in the case with smeared correlators. In addition, to be consistent 
across all $\kappa$ values, we have always determined the masses from the local
sink - smeared source ($ls$) combination. This is also true for the rho mass
where we investigated only the VV correlator.
   
 For the smeared correlators, we first determine the optimum smearing parameter
$s_0$ for each correlator PP, AA, AP and PA and in general the optimum values 
are different for different operators (see Table \ref{r0}). As
mentioned already above, for a given 
operator, the value of the optimum $s_0$ needs to be the same for all
smearing combinations $ls$, $sl$ and $ss$ to make use of the formula
given by Eq. (\ref{ls_sl_by_ss}). We have presented all the optimum $s_0$ 
values
chosen in our analysis for each operator at each $\kappa$ in Table \ref{r0}. 
There is very little systematics or a visible trend in the values except that
usually the  PP correlator needed a bigger smearing size parameter than the 
AA, the AP or the PA for which one may decipher a rough trend of increasing
$s_0$ with increasing $\kappa$. This trend is visible also in the VV 
correlator.

\begin{figure}
\centering
\includegraphics[width=4in,clip]{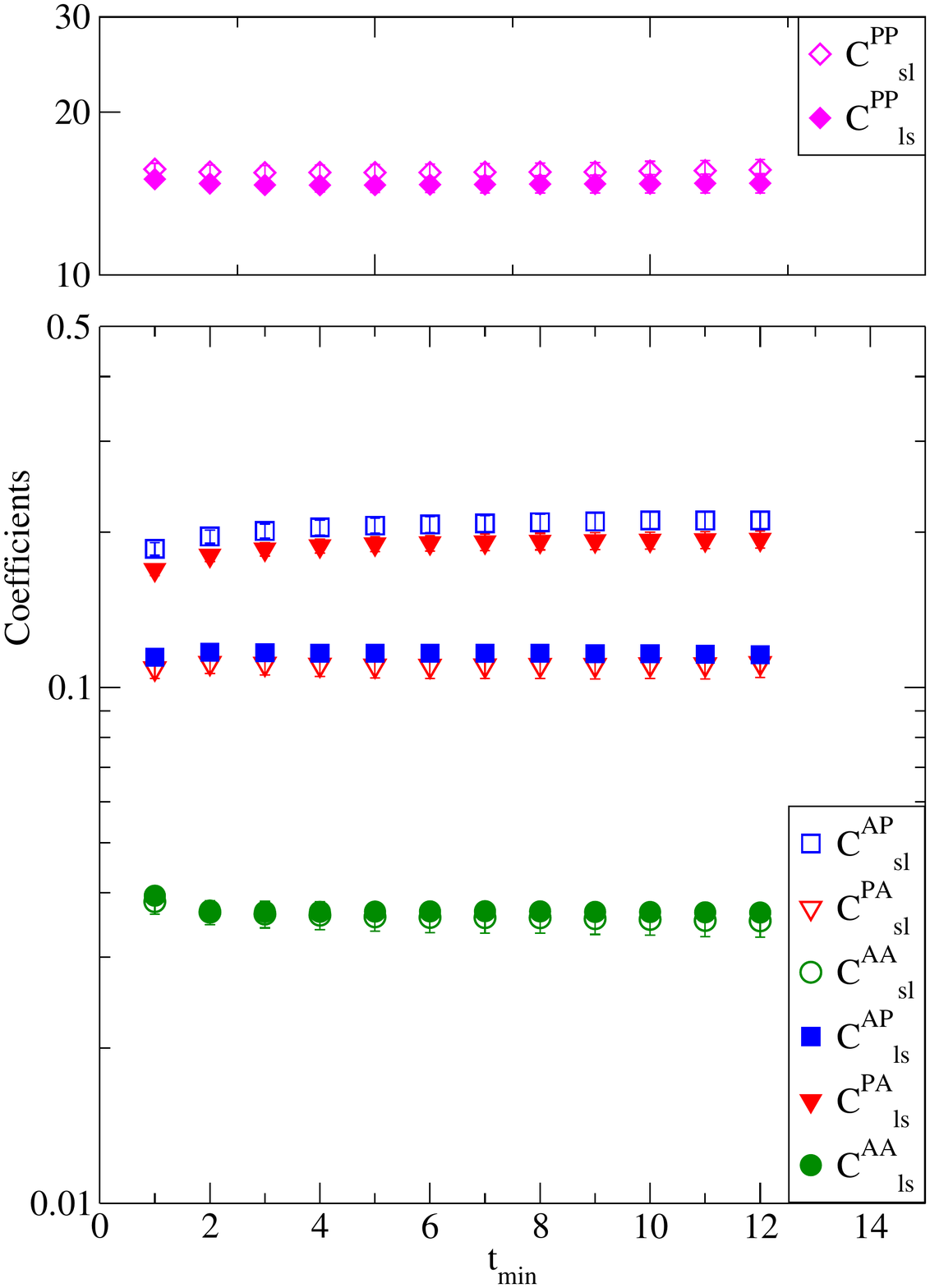}
\caption{Comparison of $ls$ and $sl$ coefficients for all the pion correlators 
at $\kappa=0.1575$.}
\label{comp_ls_sl}
\end{figure}

While the calculation of
$C_{ss}$ is mandatory, in the literature (e.g.,\cite{alikhan,orthprd}) 
$C_{sl}$ (corresponding to smeared sink and local source) has often been 
approximated by $C_{ls}$
and use $C_{ls}^2/C_{ss}$ to calculate $C_{ll}$.
Since we have performed calculations with both local sink - smeared source 
($ls$) and 
smeared source - local sink ($sl$) combinations, we can check the validity of 
this approximation . In Fig. \ref{comp_ls_sl} we plot $C_{ls}$ and $C_{sl}$ 
at $\kappa=0.1575$ for the all four pion correlators as functions of
$t_{\rm min}$.
Solid symbols are $ls$ and empty symbols are $sl$ correlators. As seen in the 
figure, all coefficients show decent plateau behavior. The data shown are 
at the respective optimum values of the smearing size parameter $s_0$ 
for each operator
($s_0=6$ for PP, $s_0=3$ for AA, AP and PA). While for the PP and the AA 
correlators, one should compare their respective $ls$ and $sl$ correlators, 
for the asymmetric correlators one should compare $C^{AP}_{ls}$ with 
$C^{PA}_{sl}$ and $C^{AP}_{sl}$ with $C^{PA}_{ls}$. As the figure also 
shows, it is grossly wrong to approximate $C^{AP}_{ls}$ by $C^{AP}_{sl}$ and 
similarly for PA. Actually the difference between $C^{AP}_{ls}$ and 
$C^{AP}_{sl}$
or between $C^{PA}_{ls}$ and $C^{PA}_{lsl}$ is significant, it is smearing 
size  dependent and in this case it is about 50\%. However, when one compares 
the 
correct quantities, i.e., $C^{AP}_{ls}$ and $C^{PA}_{sl}$, their central values
differ by about 5\%, while the central values of $C^{AP}_{sl}$ 
and $C^{PA}_{ls}$
differ by almost 10\%. The difference between $ls$ and $sl$ coefficients for 
the case of the AA is the least (about 3\%) while for PP it is about 5\%.

\begin{figure}
\centering
\includegraphics[width=6in,clip]{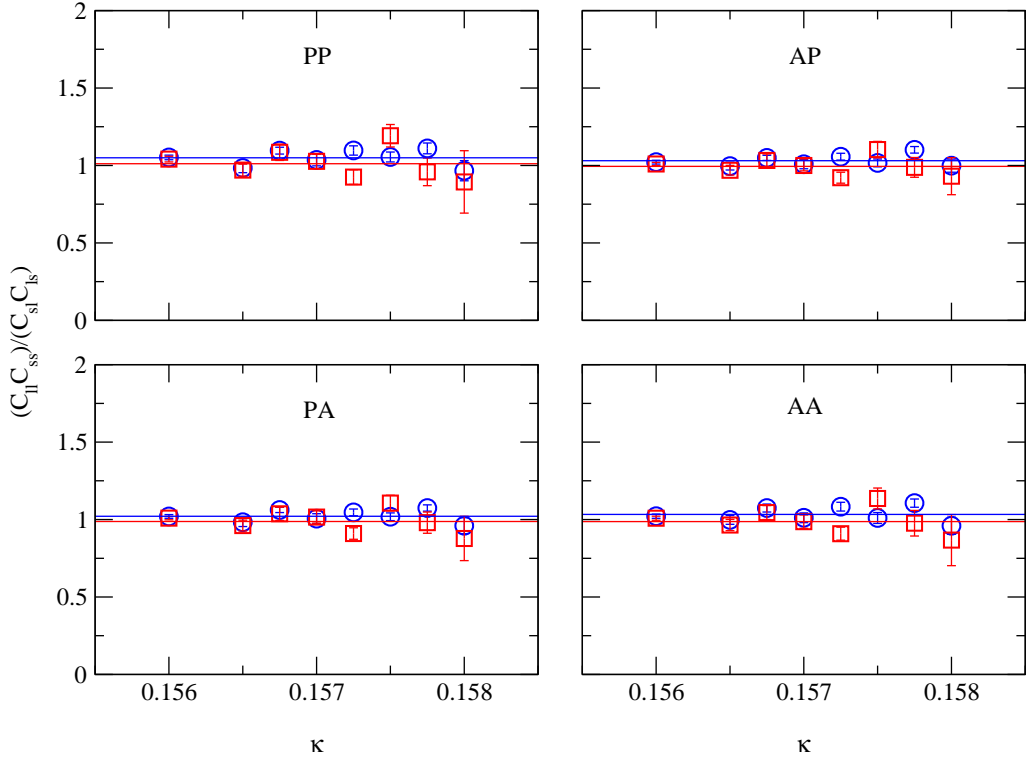}
\caption{$C_{ll}C_{ss}/C_{sl}C_{ls}$ for all four pion operators as function of
$\kappa$. This ratio should be equal to unity. The blue circles (red squares) 
represent data where the pion mass is determined from the PP (AA) correlator.}
\label{ratio_coeff}
\end{figure}

In Fig. \ref{ratio_coeff} we show $C_{ll}C_{ss}/C_{sl}C_{ls}$ for all four 
pion correlators as function of $\kappa$. This ratio should be equal to unity. 
The circles (squares) 
represent data where the pion mass was determined from the PP (AA) correlator.
The straight lines represent average values of the respective 
ratios. They are very close to unity (within a few percent). While the 
squares (pion mass taken from AA) show more fluctuation, the average is almost 
unity; the circles (mass taken from PP) are more stable, but their 
average 
shows a few percent bias above unity in all the four correlators. This general
character is displayed in all the results of the derived quantities we have 
obtained, viz., data where $m_{\pi}$ was determined from AA show more 
fluctuation than the data where $m_{\pi}$ was determined from PP, but show the 
correct general trend, in particular seem to show less finite size
effect (discussed later).
 
\begin{figure}
\centering
\includegraphics[width=4in,clip]{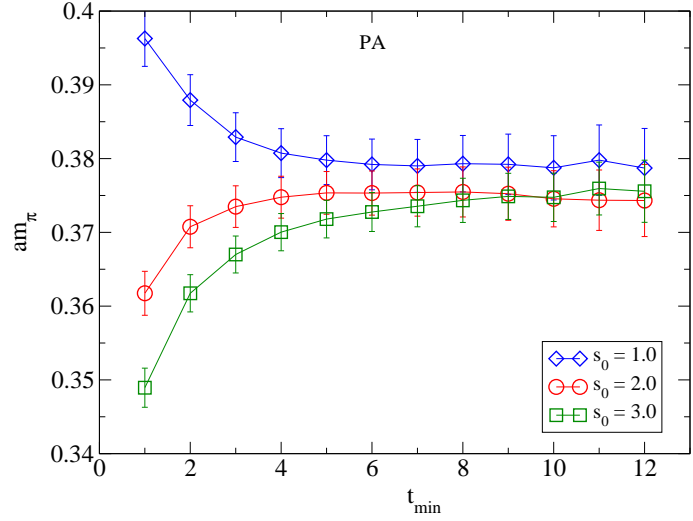}
\caption{Example of oversmearing for the $ls$ PA correlator at 
$\kappa=0.15675$.
The lines are just to guide the eye.}
\label{oversmear}
\end{figure}

Before we end the discussion on the details of the implementation of the 
gaussian smearing, we present two figures to illustrate two particular features
of the smearing. In Fig. \ref{oversmear} we show an example 
of oversmearing for 
the $ls$ PA correlator at $\kappa=0.15675$. While $s_0=1$ data look slightly 
undersmeared (albeit producing a decent plateau), $s_0=3$ completely destroys 
the plateau. $s_0=2$ produces a very stable plateau and has been accepted as 
the optimum value for this correlator, but the best value could have been 
somewhere between $s_0=1$ and $2$ (also please notice 
the significant difference
between the $m_{\pi}$ values taken from the plateaux at $s_0=1$ and $2$). Our 
choice of $s_0=2$ for this correlator at this $\kappa$ is 
also dependent on the 
proper behavior (i.e. existence of stable plateau) of the other two smearing
combinations, viz., $sl$ and $ss$ of the PA correlator at $\kappa=0.15675$.

\begin{figure}
\centering
\includegraphics[width=4in,clip]{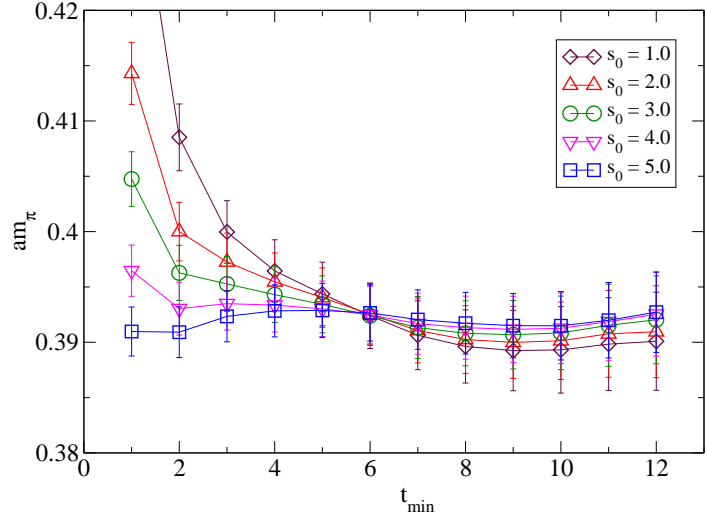}
\caption{Crossing of $am_{\pi}$ determined from the AA correlator as function 
of $t_{\rm min}$ for different smearing size parameter $s_0$ at $\kappa=0.1565$.
The lines are just to guide the eye.}
\label{crossing}
\end{figure}

Fig. \ref{crossing} plots $am_{\pi}$ (computed from $ls$ AA correlator) versus 
$t_{\rm min}$ at $\kappa=0.1565$ for $s_0=1$ to $5$. 
The curious thing to observe in
the plot is the crossing of the lines approximately at $t_{\rm min}=6$ indicating that $am_{\pi}$ is independent of $s_0$ 
for the fits using $t_{\rm min}=6$. One may 
conclude that the higher state contributions 
to the correlator are nearly absent
at the crossing point and hence the value of $am_{\pi}$ at the crossing is 
fairly accurate. This kind of crossing does not take place always, but can be 
used whenever it does. Such a behavior was observed and discussed in the 
context of static quark potential with gauge field smearing 
in Ref. \cite{ape}.

The analysis for the rho correlator VV is done in a similar way, except that
it was less tedious because VV was the only correlator we considered.

\section{Results for the pion: $am_{\pi}$, $aF_\pi$ and $am_q$}
\label{results_pi}

We present in Table \ref{piondata} all the results related to the
pion. 
The results are presented in units of the lattice constant $a$. 
The results are put in two groups depending on whether $am_\pi$ is 
determined from the PP or the AA correlator, as described in the above 
section (Sec. \ref{smearing_details}). In each group, results have
been 
presented for both the local (unsmeared) correlators and the smeared 
correlators. In each of these cases, as explained in Sec
\ref{obs_local}, 
there are three ways each to evaluate $aF_\pi$ and the PCAC quark mass $am_q$. 

\begin{table}
\begin{tabular}{|c|c|c|l|l|l|l|}
\hline \hline
$\kappa$ &\multicolumn{2}{c|}{Observables}&\multicolumn{2}{c|}
{$am_\pi$ from PP}&
 \multicolumn{2}{c|}{$am_\pi$ from AA}\\
\cline{4-7}
& \multicolumn{2}{l|}{} & {with smearing} & {without smearing} & {with
smearing} & {without smearing}\\
\hline
\hline
\hline
&  \multicolumn{2}{c|}{$am_\pi$} &{0.4503(16)} &{0.4522(19)}
&{0.4506(28)} &{0.4513(32)}\\
\cline{2-7}
&  & {$AA$} & {0.06698(87)}&{0.06628(87)} &{0.06711(83)} &
{0.06615(82)}\\
\cline{3-7}{} & {$am_q$}& {$AP$} & {0.06744(40)} & {0.06574(30)}& 
{0.06764(52)} & {0.06570(43)}\\
\cline{3-7}{0.156} & & {$PA$} & {0.06855(159)} & {0.06672(156)}& 
{0.06873(161)} & {0.06672(163)}\\
\cline{2-7}
& {} & {$AA$} & {0.09029(117)}& {0.09107(124)}&{0.09037(125)}
&{0.09068(126)} \\
\cline{3-7} &{$aF_\pi$} & {$AP$} & {0.09090(91)}& {0.09033(90)}& {0.09109(131)}
&{0.09006(108)} \\
\cline{3-7}{} & & {$PA$} & {0.09239(202)} & {0.09168(200)}& {0.09256(207)} &
{0.09145(196)}\\
\hline
\hline 
&  \multicolumn{2}{c|}{$am_\pi$} &{0.3974(17)} &{0.3956(26)} &{0.3929(24)}
&{0.3895(36)}\\
\cline{2-7}
& {} & {$AA$} & {0.05212(77)}& {0.05224(79)}& {0.05205(79)}&
{0.05145(80)}\\
\cline{3-7}{} &{$am_q$} & {$AP$} & {0.05251(46)}& {0.05293(27)}& {0.05307(43)}
&{0.05257(29)} \\
\cline{3-7}{0.1565} & & {$PA$} & {0.05146(139)} & {0.05117(147)}&
{0.05179(158)} & {0.05087(143)}\\
\cline{2-7}
&  & {$AA$} & {0.07800(117)}& {0.07809(139)}& {0.07687(122)}&
{0.07584(145)} \\
\cline{3-7} &{$aF_\pi$} & {$AP$} & {0.07859(97)}& {0.07913(93)}& {0.07838(80)}
&{0.07748(104)} \\
\cline{3-7}{} & & {$PA$} & {0.07701(194)} & {0.07650(222)}& {0.07650(217)} &
{0.07498(217)}\\
\hline
\hline
&  \multicolumn{2}{c|}{$am_\pi$} &{0.3735(17)} &{0.3780(20)}
&{0.3746(27)} &{0.3781(44)}\\
\cline{2-7}
& {} & {$AA$} & {0.04655(93)}& {0.04665(83)}& {0.04710(80)}& {0.04666(82)}\\
\cline{3-7}{} & {$am_q$}& {$AP$} & {0.04810(36)}& {0.04661(29)}& {0.04844(42)}
&{0.04655(34)} \\
\cline{3-7}{0.15675} & & {$PA$} & {0.04725(150)} & {0.04635(139)}&
{0.04763(146)} & {0.04597(148)}\\
\cline{2-7}
&  & {$AA$} & {0.07486(150)}& {0.07713(150)}&
{0.07583(130)}&{0.07716(174)} \\
\cline{3-7} &{$aF_\pi$} & {$AP$} & {0.07735(97)}& {0.07706(105)}& 
{0.07799(106)} & {0.07697(140)} \\
\cline{3-7}{} & & {$PA$} & {0.07598(229)} & {0.07662(223)}& {0.07669(223)} &
{0.07601(259)}\\
\hline
\hline
&  \multicolumn{2}{c|}{$am_\pi$} &{0.3457(18)} &{0.3456(27)}
&{0.3439(32)} &{0.3426(50)}\\
\cline{2-7}
& {} & {$AA$} & {0.04056(77)}& {0.04006(83)}& {0.04060(76)}&
{0.03963(82)}\\
\cline{3-7}{} &{$am_q$} & {$AP$} & {0.04118(40)}& {0.04007(28)}& {0.04129(55)}
&{0.04006(28)} \\
\cline{3-7}{0.157} & & {$PA$} & {0.04052(153)} & {0.03933(152)}&
{0.04044(161)} & {0.03986(135)}\\
\cline{2-7}
& {} & {$AA$} & {0.07138(137)}& {0.07179(165)}&
{0.07109(140)}&{0.07074(186)} \\
\cline{3-7} & {$aF_\pi$}& {$AP$} & {0.07249(99)}& {0.07181(97)}& {0.07230(130)}
&{0.07150(111)} \\
\cline{3-7}{} & & {$PA$} & {0.07131(258)} & {0.07049(275)}& {0.07082(280)} &
{0.07114(253)}\\
\hline
\hline
&  \multicolumn{2}{c|}{$am_\pi$} &{0.3147(20)} &{0.3194(25)}
&{0.3122(26)} &{0.3030(48)}\\
\cline{2-7}
& {} & {$AA$} & {0.03491(64)}& {0.03520(70)}& {0.03500(65)}&
{0.03367(75)}\\
\cline{3-7}{} & {$am_q$}& {$AP$} & {0.03431(33)}& {0.03355(30)}& {0.03451(36)}
&{0.03334(32)} \\
\cline{3-7}{0.15725} & & {$PA$} & {0.03632(126)} & {0.03516(122)}&
{0.03656(127)} & {0.03490(127)}\\
\cline{2-7}
& {} & {$AA$} & {0.06762(129)}& {0.06986(157)}&
{0.06731(129)}&{0.06513(170)} \\
\cline{3-7} &{$aF_\pi$} &{$AP$} & {0.06648(86)}& {0.06658(90)}& {0.06636(93)}
&{0.06450(99)} \\
\cline{3-7}{} & & {$PA$} & {0.07037(237)} & {0.06979(245)}& {0.07032(233)} &
{0.06751(256)}\\
\hline     
\hline
&  \multicolumn{2}{c|}{$am_\pi$} &{0.2876(20)} &{0.2890(29)}
&{0.2893(32)} &{0.3000(61)}\\
\cline{2-7}
& {} & {$AA$} & {0.02663(81)}& {0.02618(85)}& {0.02706(77)}&
{0.02740(78)}\\
\cline{3-7}{} &{$am_q$} & {$AP$} & {0.02896(36)}& {0.02804(30)}& {0.02933(61)}
&{0.02814(33)} \\
\cline{3-7}{0.1575} & & {$PA$} & {0.02697(125)} & {0.02615(122)}&
{0.02731(136)} & {0.02630(112)}\\
\cline{2-7}
& {} & {$AA$} & {0.06073(197)}& {0.06089(216)}&
{0.06153(176)}&{0.06440(202)} \\
\cline{3-7} & {$aF_\pi$}& {$AP$} & {0.06604(120)}& {0.06521(122)}& 
{0.06670(146)} & {0.06613(125)} \\
\cline{3-7}{} & & {$PA$} & {0.06150(288)} & {0.06082(294)}& {0.06212(301)} &
{0.06181(271)}\\
\hline
\hline
 &  \multicolumn{2}{c|}{$am_\pi$} &{0.2481(38)} &{0.2554(39)}
&{0.2360(49)} &{0.2345(76)}\\
\cline{2-7}
& {} & {$AA$} & {0.02173(58)}& {0.02233(60)}& {0.02116(70)}&
{0.02122(61)}\\
\cline{3-7}{} & {$am_q$}& {$AP$} & {0.02039(36)}& {0.02081(28)}& {0.02116(42)} 
&{0.02167(42)} \\
\cline{3-7}{0.15775} & & {$PA$} & {0.02298(94)} & {0.02289(100)}&
{0.02270(126)} & {0.02306(124)}\\
\cline{2-7}
& {} & {$AA$} & {0.05523(162)}& {0.05726(169)}&
{0.05325(189)}&{0.05279(187)} \\
\cline{3-7} &{$aF_\pi$} & {$AP$} & {0.05182(100)}& {0.05337(91)}& {0.05327(99)}
&{0.05392(109)} \\
\cline{3-7}{} & & {$PA$} & {0.05840(240)} & {0.05871(268)}& {0.05714(337)} &
{0.05737(329)}\\
\hline
\hline
&  \multicolumn{2}{c|}{$am_\pi$} &{0.2278(28)} &{0.2255(53)}
&{0.2142(81)} &{0.2051((167)}\\
\cline{2-7}
& {} & {$AA$} & {0.01481(67)}& {0.01462(69)}& {0.01452(75)}&
{0.01369(81)}\\
\cline{3-7}{} & {$am_q$}& {$AP$} & {0.01348(44)}& {0.01380(44)}& {0.01435(62)}
&{0.01431(52)} \\
\cline{3-7}{0.158} & & {$PA$} & {0.01592(122)} & {0.01566(125)}&
{0.01664(119)} & {0.01564(128)}\\
\cline{2-7}
& {} & {$AA$} & {0.04658(222)}& {0.04587(225)}&
{0.04454(244)}&{0.04239(261)} \\
\cline{3-7} & {$af_\pi$}& {$AP$} & {0.04239(141)}& {0.04330(148)}&
 {0.04404(166)} 
&{0.04432(170)} \\
\cline{3-7}{} & & {$PA$} & {0.05006(395)} & {0.04915(401)}& {0.05105(365)} &
{0.04843(406)}\\
\hline\hline
\end{tabular}
\caption{Results for $am_\pi$, $aF_\pi$ and PCAC quark mass $am_q$.}\label{piondata}
\end{table}
\vskip .2in

\begin{figure}
\centering
\includegraphics[width=6in,clip]{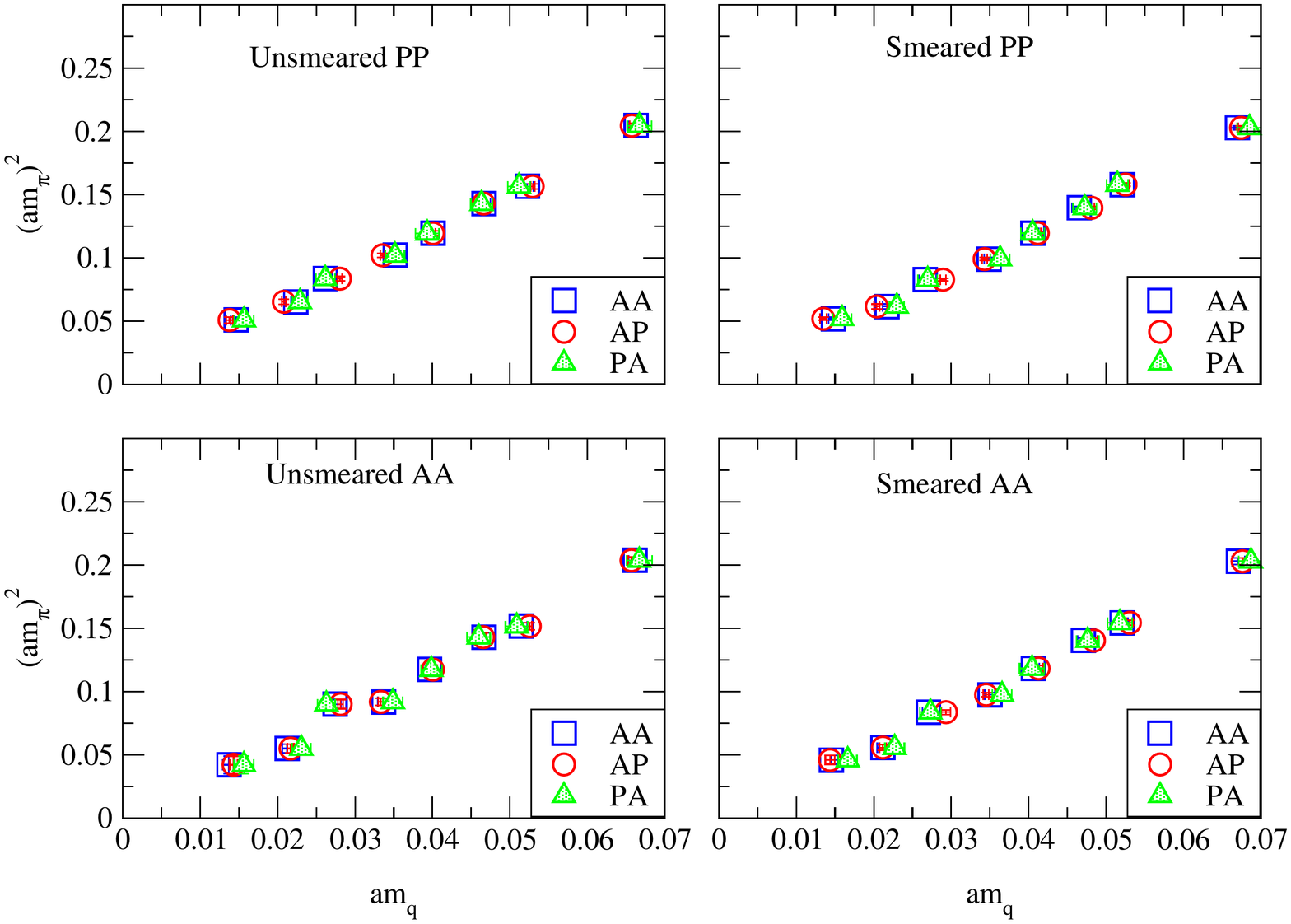}
\caption{$(am_{\pi})^2$ obtained from unsmeared and smeared PP and AA 
correlators versus different evaluations of $am_q$.}
\label{ampisq_vs_amq}
\end{figure}

\begin{figure}
\centering
\includegraphics[width=4in,clip]{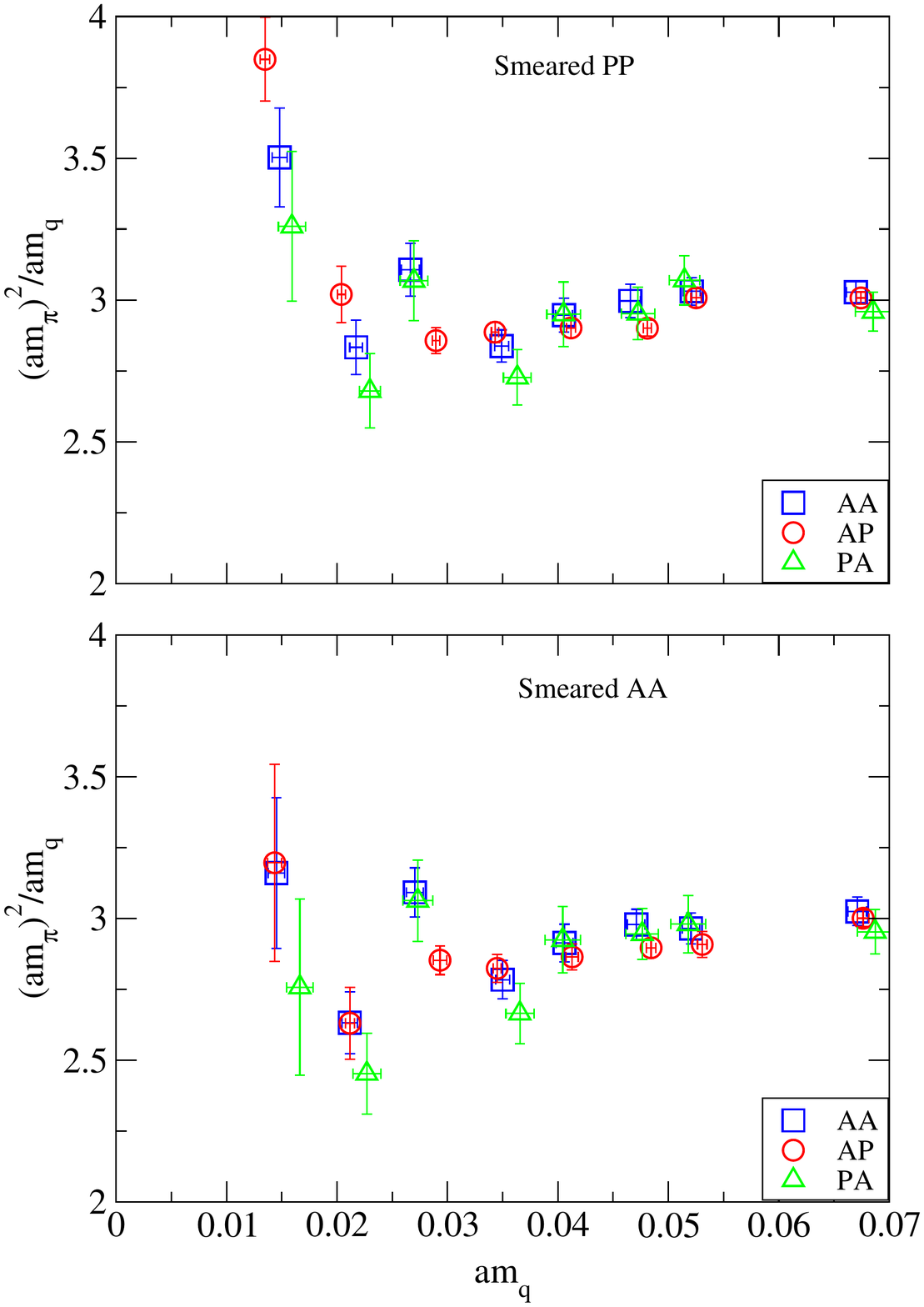}
\caption{The ratio $(am_{\pi})^2/(am_q)$ obtained from smeared 
PP and AA correlators versus different evaluations of $am_q$.}
\label{chiralratio}
\end{figure}

The errors presented in Table \ref{piondata} are statistical errors calculated by the single-omission jackknife method calculated from 200 jackknife bins. In general, the $am_\pi$ errors increase as $\kappa$ increases. These errors are systematically more for the pion masses determined from the AA correlator than from the PP correlator and within each group more for the determination from the unsmeared correlators than the smeared correlators. $aF_\pi$ is the most inaccurate quantity presented in Table \ref{piondata}. In each group, both $aF_\pi$ and $am_q$ have the least errors when the correlator AP is used; they are the most inaccurate when the correlator PA is used. For determination of quantities when pion mass was determined from the unsmeared AA correlator, at times the quality of the plateau was poor resulting in inaccurate data; this was especially true for $\kappa=0.15725$ and $0.1575$. 

Fig. \ref{ampisq_vs_amq} plots $(am_{\pi})^2$ determined from
unsmeared and smeared PP and AA correlators versus different
evaluations of $am_q$.  Except for the data points with the lowest
pion and quark masses, the rest appears to be roughly consistent with
the lowest
order (LO) chiral perturbation theory ($\chi PT$). As already
mentioned, out of 
the four plots in Fig.  \ref{ampisq_vs_amq}, the bottom left plot 
(pion mass from unsmeared AA) looks the worst. 
In all cases, the PCAC quark mass $am_q$ determined from the 
AP and the PP correlator (shown in plots as AP: open circles) 
has the least error.

Any departure from LO $\chi PT$ shows up in the behavior of the ratio
$(am_{\pi})^2/(am_q)$, called the {\em chiral ratio} below, 
as a function $am_q$. This is plotted for the
data obtained only from the smeared correlators in
Fig. \ref{chiralratio}.  Firstly we need to separate  any effect of
finite size effect on the masses before discussing the chiral
behavior. We are quite confident that for the values of $\kappa$ from
0.156 to 0.1575, there is negligible finite size effect on these
masses, because
data, consistent with our numbers, are available in previous
literature at some of these $\kappa$ values at larger
volumes (see below for discussion and references), 
For the masses at $\kappa=0.15775$, our guess is that for different operators there is a difference in finite size effect. From both the Figs. \ref{ampisq_vs_amq} and \ref{chiralratio}, it seems that there is a little bit of finite size effect for the pion mass at this $\kappa$, but more with pion mass determined from PP than AA, because in the upper plot of 
Fig. \ref{chiralratio} there is an upward trend at this point. This is
not to be taken as an effect from next-to-leading order (NLO) $\chi
PT$ because with standard values of low energy effective (LE)
constants of $\chi PT$ obtained from other studies (both on the
lattice and otherwise), there should not be an upward trend of the ratio (as seen in the upper plot here) at this pion mass. 

Difference in the finite size effect on the pion mass calculated from different operators is an interesting observation and to the best of our knowledge has not been discussed in the literature before. It is fair to say that this is not an unexpected behavior. However, what is also interesting is that  the generally accepted noisier operator AA produces less finite size effect.

If our data see only the LO $\chi PT$, then  the ratio should be a
perfect straight line parallel to the $am_q$ axis. If we believe our
error bars, there is a significant downward trend of the ratio from
the larger quark masses to the smaller quark masses (more pronounced
with the AA and PA quark masses). For the lower plot of Fig. \ref{chiralratio} this trend continues and intensifies to the second last point from the left (at $\kappa=0.15775$) which can only be an effect from higher order terms of $\chi PT$.

The left-most point (at $\kappa=0.158$) definitely has a significant
finite size effect as we can compare our data with evaluations on
larger volumes (see below). Even at this point, the pion mass determined from the smeared AA correlator has smaller finite size effect as apparent from the figure.  We also like to point out that the finite size effects, its dependence on the pion operators etc have nothing to do with smearing or the smearing size \cite{milcw}. The same effects are also visible from similar data from unsmeared operators (although not plotted in Fig. \ref{chiralratio}, but available in Table \ref{piondata} and Fig. \ref{ampisq_vs_amq}).    

\begin{figure}
\centering
\includegraphics[width=6in,clip]{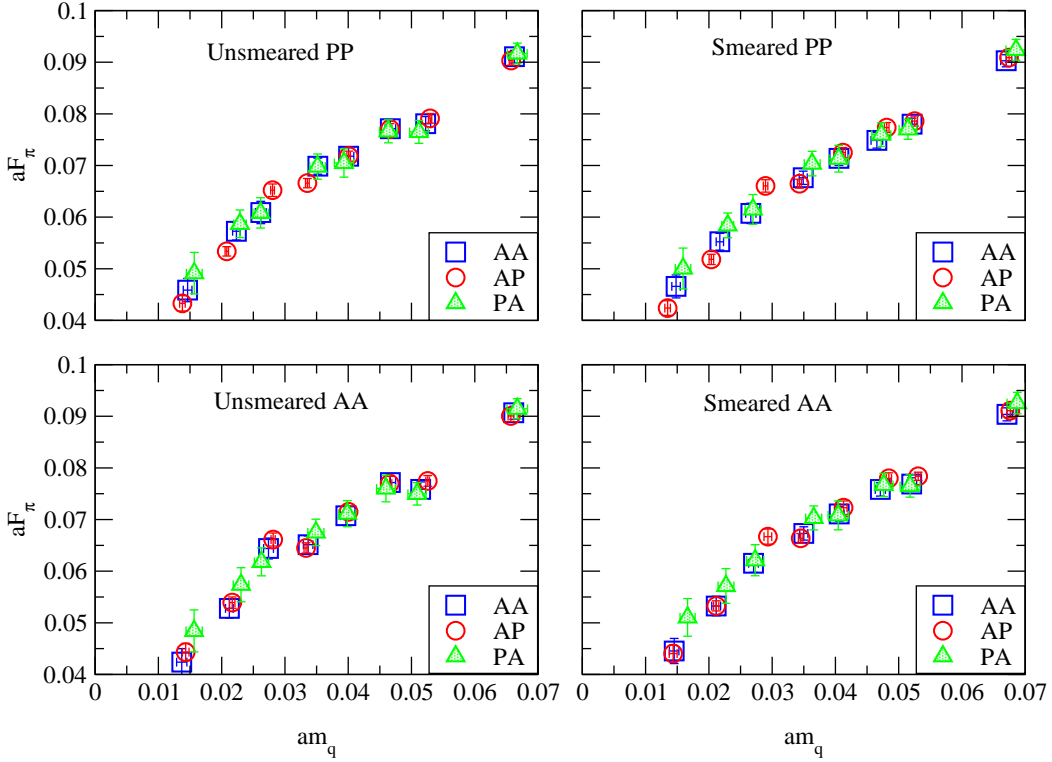}
\caption{$aF_{\pi}$ obtained from unsmeared and smeared PP and AA 
correlators versus different evaluations of $am_q$.}
\label{aFpi_vs_amq}
\end{figure}

Fig. \ref{aFpi_vs_amq} plots $aF_\pi$ versus $am_q$ in way similar to
Fig. \ref{ampisq_vs_amq}. The bottom left plot where the pion mass is
determined from unsmeared AA correlator has the most inaccuracies. The
main difference with the $am_\pi^2$ plot is that the finite size
effect at $\kappa=0.15775$ and 0.158 is now quite severe. This is
consistent with $\chi PT$ which predicts four times larger (and
opposite in sign) finite size effect for $aF_\pi$ as compared to
$am_\pi^2$ \cite{cola}.  The behavior of $aF_\pi$ with $am_q$ using
the AA operator (squares) show the most continuous behavior, 
even though these have larger error-bars compared to the 
$am_q$ data computed from the AP correlator.   

\subsection{Chiral behavior}
\begin{figure}
\centering
\includegraphics[width=6in,clip]{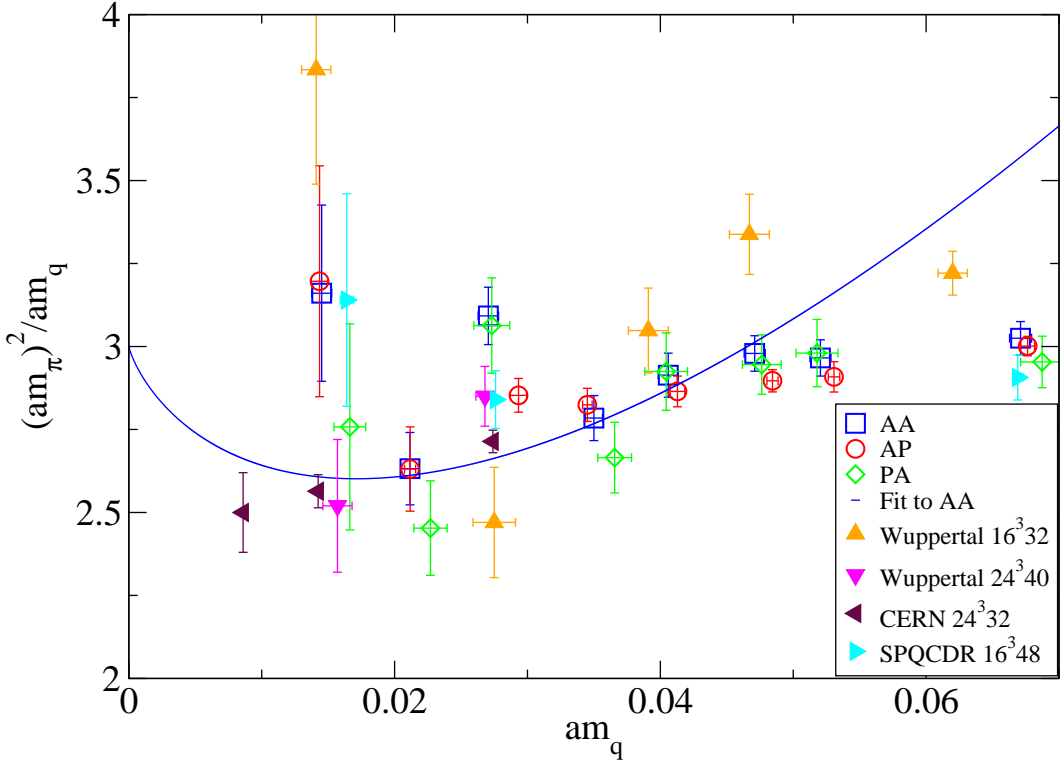}
\caption{The chiral ratio $(am_\pi)^2/am_q$ obtained from smeared AA 
correlator versus different evaluations of $am_q$. Figure includes
data from other work at same or larger volume and a suggestive NLO chiral plot.}
\label{MsqPi_by_mQ_chiral}
\end{figure}

\begin{figure}
\centering
\includegraphics[width=6in,clip]{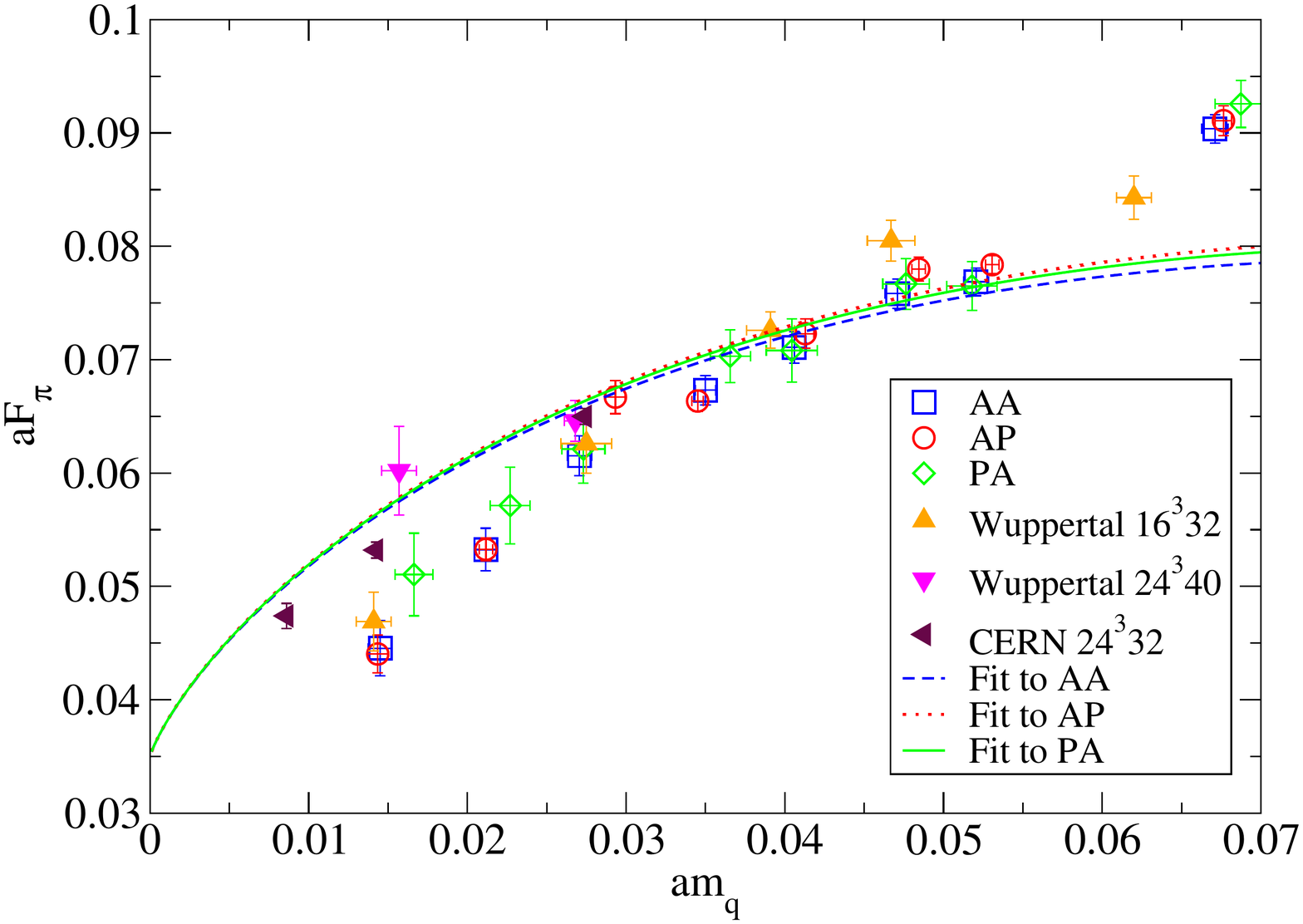}
\caption{$aF_{\pi}$ obtained from smeared AA 
correlator versus different evaluations of $am_q$. Figure includes
data from other work at same or larger volume and a suggestive NLO chiral plot}
\label{aFpi_chiral}
\end{figure}

We have plotted in Figs. \ref{MsqPi_by_mQ_chiral} and \ref{aFpi_chiral} 
the whole content of
the bottom figures of Fig. \ref{chiralratio} and \ref{aFpi_vs_amq} respectively
(data from smeared AA
correlator only, represented by open symbols) along with (unimproved)
$N_F=2$ Wilson data at $\beta=5.6$ from Ref.
\cite{orthprd} ($16^332$ and $24^340$), Ref. \cite{besirevic} ($16^348$) 
and Ref. \cite{deldebbio2} ($24^332$) (data from others represented by
different filled symbols). These figures also show  NLO
$\chi PT$ \cite{leut} plots 
\begin{eqnarray}
\frac{(am_\pi)^2}{am_q} &=& 2aB\left[ 1-\frac{m_qB}{16\pi^2F^2}{\rm ln}
\frac{\Lambda_3^2}{2m_qB} \right] \\
aF_\pi &=& aF \left[ 1+\frac{m_qB}{8\pi^2F^2}{\rm ln}
\frac{\Lambda_4^2}{2m_qB} \right]
\end{eqnarray} 
with $aF=0.035$, $aB=1.5$. With
these values of $aF$ and $aB$ as input we actually could fit our
data (for $am_q^{AA}$, represented by squares) for both the chiral ratio 
and $aF_\pi$. For the chiral ratio, the fits include points from 
$\kappa=0.15675, 0.157, 0.15725$ and 0.15775. 
The point at $\kappa=0.1575$ is excluded from
the fit because of systematic fluctuations.
The fit to the $aF_\pi$ data is also done with the AP and PA
quark masses in a similar range of data. The values of $a\Lambda_3$ and
$a\Lambda_4$ that come out of the fits are: $a\Lambda_3= 0.374(7)$, 
$a\Lambda_4=0.813(11)$. 

Let us state categorically that the chiral NLO plots are not to be
taken as serious fits, they are shown more to understand our data and
where the data of other works (with same parameters), especially on
larger volumes, are in relation to our data and the chiral NLO
plots. Firstly, the NLO fits for both the chiral ratio and the
$aF_\pi$ are done with the same input values of $aF$ and
$aB$. Secondly, the fit to the chiral ratio is done with $am_q^{AA}$
(denoted by open squares) and nearly goes through all the points of
the fit, Our other data points with $am_q^{AP}$ (circles, with their 
small error bars) and with $am_q^{PA}$ (diamonds, with large error
bars) are also close to this NLO curve. Thirdly, the data points on
larger lattices (at $\kappa=0.158$ by \cite{deldebbio2, orthprd}, at
$\kappa=0.1575$ by \cite{deldebbio2} lie very close to both the NLO
fits (Figs. \ref{MsqPi_by_mQ_chiral} and \ref{aFpi_chiral}). The data
with the smallest quark mass from \cite{deldebbio2} 
at $\kappa=0.15825$ has a large error
bar on the chiral ratio plot, however, on the $aF_\pi$ plot looks very
close to the NLO fit. The errors for the $16^3$ data from 
\cite{orthprd,besirevic} are calculated by us from their quoted pion and
quark mass data by quadrature (Ref. \cite{besirevic} has no $F_\pi$
data either). Some of these $16^3$ data look a bit erratic,
especially from \cite{orthprd}. Finally, probably because of large
finite size effect on $aF_\pi$, the numerical data computed on
relatively smaller lattices (including ours) tend to have a larger
downward bend than is consistent with acceptable values for the low
energy constants. However, our fits with $aF$ and $aB$ as inputs, done
from $\kappa=0.1565$ to 0.1575, for all the quark masses $am_q^{AA}$,
$am_q^{AP}$ and $am_q^{PA}$ give a reasonable fit with the evaluations
upto $\kappa=0.15825$ on larger lattices staying very close to this
explorative NLO fit. 

At this point if we look back at the Fig. \ref{ampisq_vs_amq} and take
a closer look, we discover that this figure is {\em not} consistent
with LO $\chi PT$, because data at larger masses do {\em not} go
though the origin in a straight line. The data in all the plots of 
Fig. \ref{ampisq_vs_amq} actually have a small bend.

\subsection{Comparison of VWI and AWI masses}

In all the Tables and Figures above, we have presented results for
the quark mass from the PCAC relation, $am_q^{AWI}$, which is
devoid of ${\cal O}(a)$ effects because the PCAC relation guarantees
that the square of the pion mass vanishes as the quark mass approaches
zero. $\chi PT$ can also be made consistent with this, once the quark
masses used there are the PCAC masses \cite{ss}. 

Traditionally, one also defines the VWI (vector Ward Identity) quark
mass as
\begin{eqnarray}
am^{VWI}_{q} = \frac{1}{2}\left ( \frac{1}{\kappa} -
\frac{1}{\kappa_c} 
\right), \label{VWI}
\end{eqnarray}
where $\kappa\rightarrow\kappa_c=1/8=0.125$ in the chiral limit for the
free theory ($U_{x\mu}=1$).

At the tree level, this definition takes care of the ${\cal O}(a)$
additive quark mass. However, at the quantum level, $m^{VWI}_{q}$
should be devoid of ${\cal O}(a)$ effects only approximately. The
effect should be worse at smaller values of $\beta$ with a larger
lattice constant $a$.  

\begin{figure}
\subfigure{
\includegraphics[width=3in,clip]{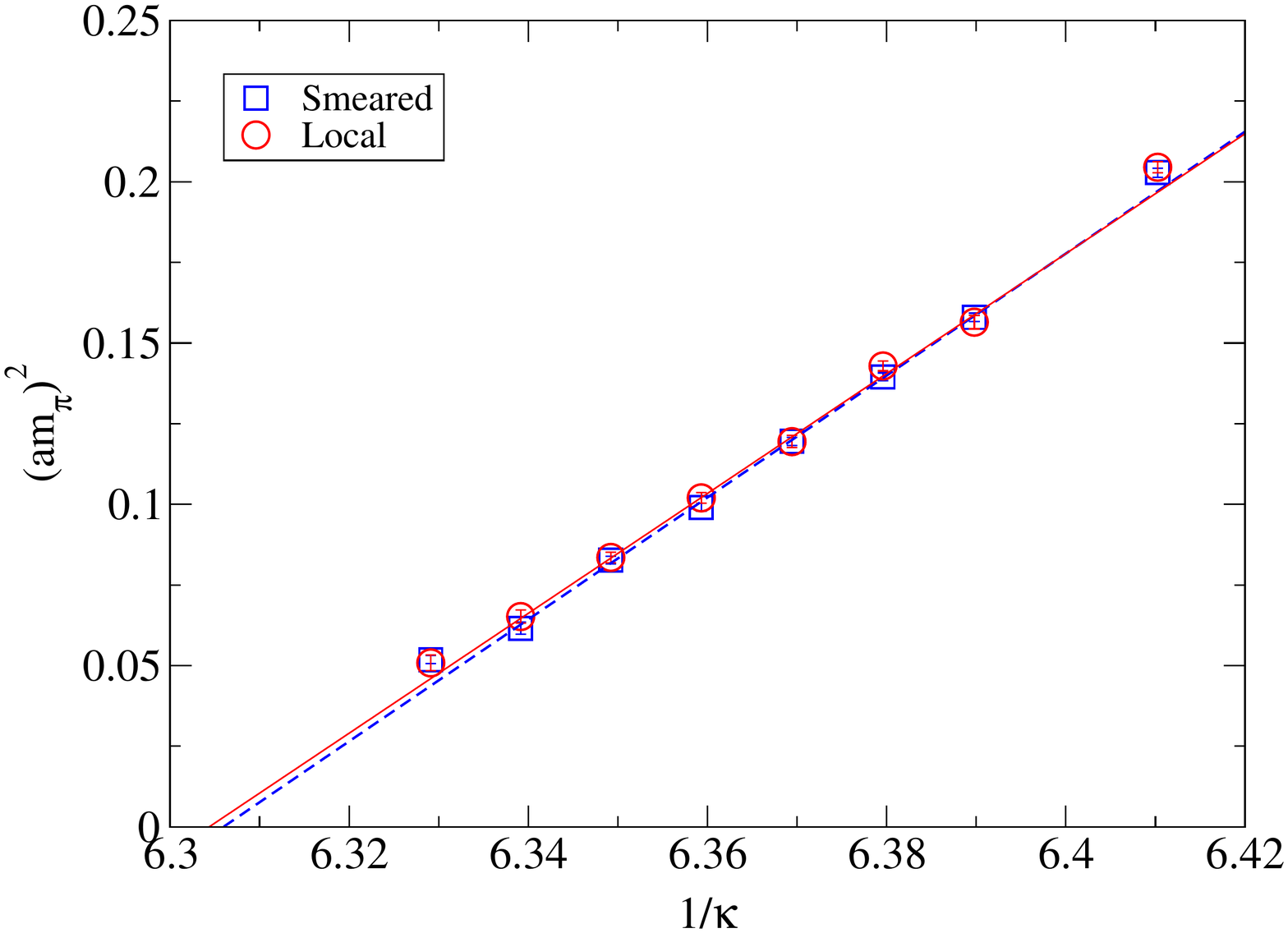}
}
\subfigure{
\includegraphics[width=3in,clip]{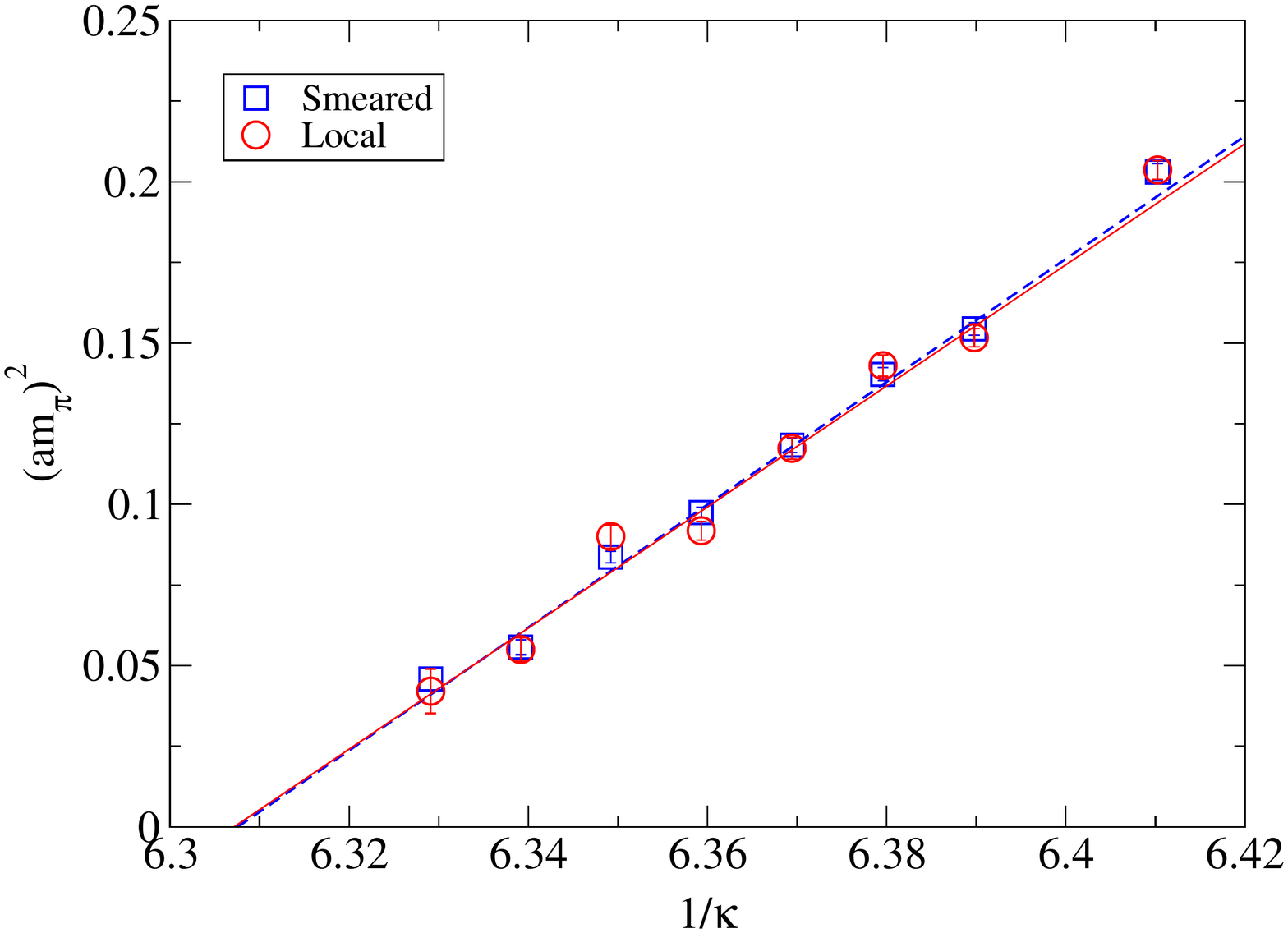}
}
\caption{$(am_{\pi})^2$ obtained from unsmeared and smeared PP (left) and 
AA (right) correlators versus $\frac{1}{\kappa}$.}
\label{kappa_c}
\end{figure}

The behaviour of $(am_\pi)^2$ as function of $\frac{1}{\kappa}$
is shown in Fig. \ref{kappa_c} for pion masses determined from
unsmeared and smeared PP and AA correlators. Although the data from
unsmeared AA correlators show more fluctuations, straight lines fits
excluding $\kappa=0.158$ (because of sizable finite size effects)
are possible in all cases giving a determination of $\kappa_c$. The
values of $\kappa_c$ obtained are as follows: (i) 0.15858(3)
(PP-smeared), (ii) 0.15862(4) (PP-unsmeared), (iii) 0.15854(4)
(AA-smeared), and (iv) 0.15855(6) (AA-unsmeared). For the same lattice
volume and $\beta$, SESAME and T$\chi$L Collaboration \cite{tlqcd} obtained 
$\kappa_c=0.158493(16)$.

The values  of $am^{VWI}_q$ calculated using Eq. (\ref{VWI})
are presented in Table \ref{mqvwi}.

\begin{table}
\begin{tabular}{|l|l|l|l|l|}
\hline \hline
$\kappa$ &\multicolumn{2}{c|}{$am_\pi$ from PP}&
 \multicolumn{2}{c|}{$am_\pi$ from AA}\\
\cline{2-5}
&  {with smearing} & {without smearing} & {with smearing} & {without smearing}\\
\hline
{0.156}&{0.05215(55)} &{0.05294(75)} &{0.05135(75)} &{0.05155(126)}\\
\hline 
{0.1565}&{0.04191(55)} &{0.04270(75)} &{0.04111(75)} &{0.04131(126)}\\
\hline
{0.15675}&{0.03681(55)} &{0.03761(75)} &{0.03601(75)} &{0.03621(126)}\\
\hline
{0.157}&{0.03173(55)} &{0.03253(75)} &{0.03094(75)} &{0.03113(126)}\\
\hline
{0.15725}&{0.02667(55)} &{0.02746(75)} &{0.02587(75)} &{0.02607(126)}\\
\hline
{0.1575}&{0.02162(55)} &{0.02242(75)} &{0.02083(75)} &{0.02102(126)}\\
\hline
{0.15775}&{0.01659(55)} &{0.01738(75)} &{0.01579(75)} &{0.01599(126)}\\
\hline
{0.158}&{0.01157(55)} &{0.01237(75)} &{0.01078(75)} &{0.01098(126)}\\
\hline\hline
\end{tabular}
\caption{$am_q^{VWI}$ in dependence of $\kappa$. }
\label{mqvwi}
\end{table}

\begin{figure}
\centering
\includegraphics[width=4in,clip]{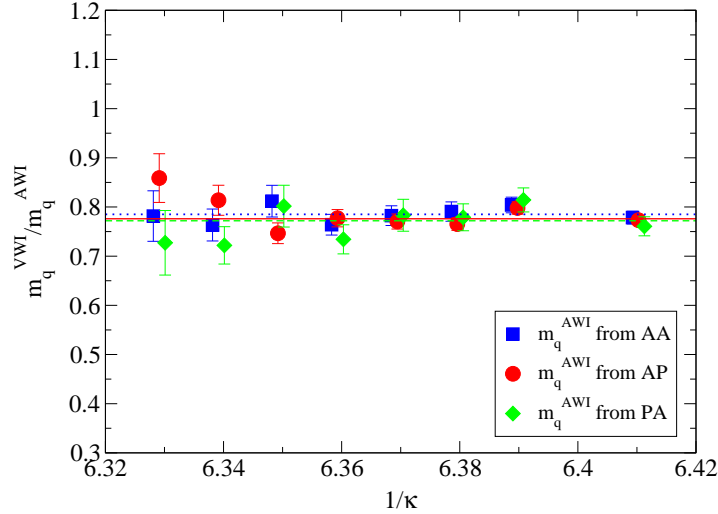}
\caption{Deviation of  of $m_q^{VWI}$ from $m_q^{AWI}$ (some of the
  data slightly shifted horizontally for clarity). }
\label{qmassratios}
\end{figure}

We note that $m_q^{AWI}$ has no ${\cal O}(a)$ scaling violation whereas
$m_q^{VWI}$ has some remaining ${\cal O}(a)$ scaling violation. 
Fig. \ref{qmassratios}
plots the deviation of 
  $m_q^{VWI}$ from $m_q^{AWI}$ as a function of $\frac{1}{\kappa}$. 
We notice small fluctuations (a few percent) around a normalization
factor of approximately
0.78 (i.e., $m_q^{AWI} \approx 0.78 m_q^{VWI}$). Even at $\kappa=0.158$
where the fluctuation is maximum shows a deviation of about 6\% from
the normalization. If these deviations are any signature, scaling
violations may be small in our case, This is to be expected at our
lattice scale $a\approx 0.08 {\rm fm}$ ($a^{-1}\sim 2.45 {\rm GeV}$) 
\cite{qcd_paper1}.   

\section{Results for the Rho: $am_\rho$, $aF_\rho$ and different ratios}
\label{results_rho}
Table \ref{rhodata} presents our results for the rho mass and the
decay constant 
in lattice units for all the $\kappa$ values of our simulation. 
The errors are again the single-omission jackknife errors computed
from 200 jackknife 
bins. In general the errors are more for larger $\kappa$  and for data 
from unsmeared correlators (especially for $am_\rho$). 
With the range of pion masses reached in our simulations, 
$2m_\pi> m_\rho$ was always satisfied and there was no complication in 
the investigation of the rho correlator. 

\begin{figure}
\centering
\includegraphics[width=4in,clip]{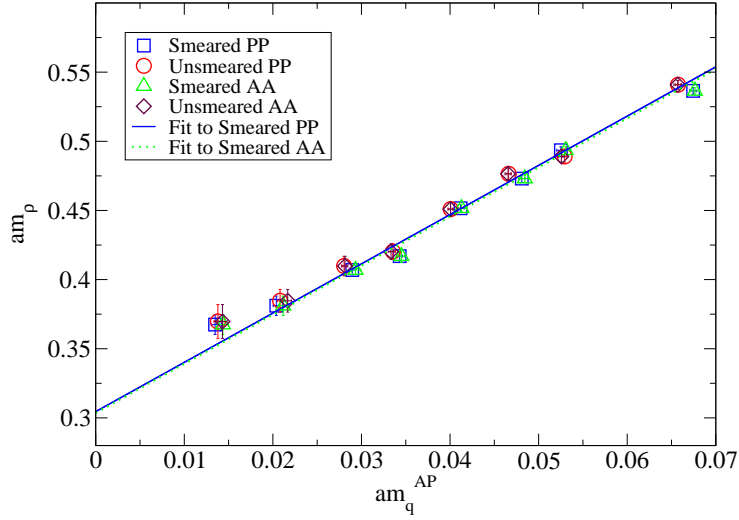}
\caption{$am_\rho$ obtained from unsmeared and smeared PP and AA 
correlators versus $am_q^{AP}$.}
\label{amrho_vs_amqAP}
\end{figure}

The data for $am_\rho$ is plotted in Fig. \ref{amrho_vs_amqAP} as a 
function of $am_q^{AP}$. For clarity of the figure, other
quark masses are not shown in this figure (even in this case, there
are four different PCAC quark masses depending on  
whether the pion mass was determined from unsmeared or smeared PP or AA
correlators). 
The straight line fits shown in the figure exclude points for the two
extreme $\kappa$ and for $\kappa=0.15725$ (appearing to have a systematic
error) and lead to a value close to the 
physical rho mass on chiral extrapolation. As expected, the lowest rho
mass at $\kappa=0.158$ has some finite size effect.

\begin{figure}
\centering
\includegraphics[width=6in,clip]{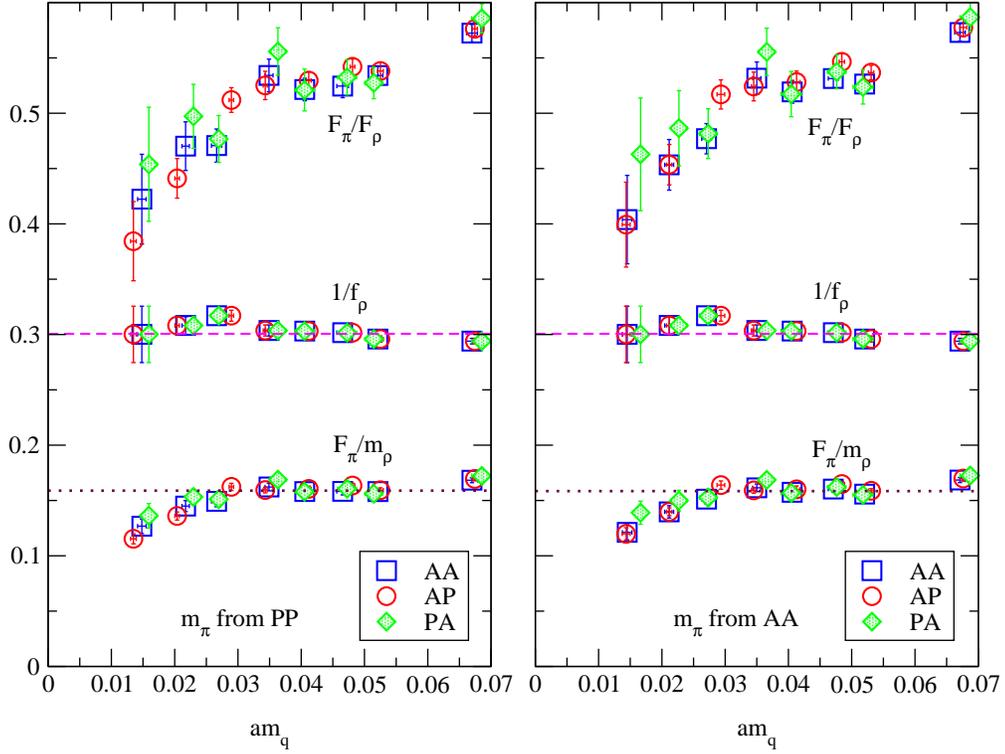}
\caption{$F_\pi/F_\rho$, $1/f_\rho=F_\rho/m_\rho$ and $F_\pi/m_\rho$ obtained 
from smeared correlators versus different evaluations of $am_q$.}
\label{rho_ratios}
\end{figure}

Fig. \ref{rho_ratios} shows three ratios $F_\pi/F_\rho$,
$1/f_\rho=F_\rho/m_\rho$ 
and $F_\pi/m_\rho$  obtained from smeared correlators versus 
PCAC quark masses. The left (right) figure shows the dependence of
these 
ratios on AA, AP and PA quark masses when the pion mass was 
determined from the smeared PP (AA) correlator. The drop at small quark
masses 
in the ratios $F_\pi/F_\rho$,  and $F_\pi/m_\rho$ are mainly 
attributable to the most significant finite size effect of $F_\pi$.  
The quantity in the middle $1/f_\rho=F_\rho/m_\rho$ is actually 
a ratio of the other two ratios and with the elimination of $F_\pi$  
cancels out whatever finite size effects of $F_\rho$ and $m_\rho$.     

With phenomenological inputs of $m_\rho=771.1 MeV$, $F_\pi=92.4 MeV$
and $\sqrt{2}F_\rho=216 MeV$, Fig. \ref{rho_ratios} also enables approximate 
estimations of the renormalization
constants $Z_V$ and $Z_A$ associated with $F_\rho$ and $F_\pi$
respectively. Fit to all points of the ratio $1/f_\rho=F_\rho/m_\rho$
leads to $Z_V=0.659(2)$ and a similar fit to points corresponding to 
$\kappa=$0.1565 to 0.15725 for the ratio $F_\pi/m_\rho$ leads to
$Z_A=0.76(1)$. These are remarkably close to the values quoted in 
\cite{deldebbio1,besirevic}.

\begin{table}
\begin{tabular}{|l|l|l|l|l|}
\hline \hline
$\kappa$ &\multicolumn{2}{c|}{$am_\rho$}&
 \multicolumn{2}{c|}{$aF_\rho$}\\
\cline{2-5}
&  {with smearing} & {without smearing} & {with smearing} & 
{without smearing}\\
\hline
{0.156}&{0.5365(23)} &{0.5409(32)} &{0.1577(17)} &{0.1631(24)}\\
\hline 
{0.1565}&{0.4935(25)} &{0.4890(41)} &{0.1460(17)} &{0.1443(27)}\\
\hline
{0.15675}&{0.4730(27)} &{0.4765(43)} &{0.1427(16)} &{0.1460(28)}\\
\hline
{0.157}&{0.4516(28)} &{0.4510(50)} &{0.1369(13)} &{0.1385(30)}\\
\hline
{0.15725}&{0.4170(45)} &{0.4202(56)} &{0.1266(31)} &{0.1300(30)}\\
\hline
{0.1575}&{0.4069(34)} &{0.4098(70)} &{0.1290(25)} &{0.1323(40)}\\
\hline
{0.15775}&{0.3811(70)} &{0.3847(82)} &{0.1175(47)} &{0.1221(44)}\\
\hline
{0.158}&{0.3675(71)} &{0.3697(122)} &{0.1103(99)} &{0.1130(62)}\\
\hline\hline
\end{tabular}
\caption{$am_\rho$ and $aF_\rho$ from smeared and unsmeared VV correlators for all the values of $\kappa$ }\label{rhodata}
\end{table}

\section{Summary and conclusions}
\label{conclu}
Due to recent theoretical, algorithmic and technological advances,
Lattice QCD with Wilson fermions has entered a very active phase.
We have planned on a detailed lattice QCD investigation with Wilson
and Wilson-type fermions. In this work we have taken the first step 
towards addressing the issues that are encountered probing the chiral 
regime of  lattice QCD systematically.  Here, our emphasis is to have 
fully dynamical simulations at an
extensive set of $\kappa$ values and study accurate determinations of
the masses and decay constants in as many ways as possible.
We employ 
standard (unimproved) Wilson gauge and fermion actions with 2 fully dynamical
light quark flavors on $16^{3}32$ lattice, $\beta=6/g^2=5.6$ at 8 values
of bare quark masses (corresponding to fermionic hopping parameter
$\kappa =0.156,~ 0.1565,~0.15675, ~0.157, ~0.15725,
~0.1575,~0.15775,~0.158$). The lattice scale reached is respectable and
$a\sim 0.08$ fm ($a^{-1}\sim 2.45$ GeV) \cite{qcd_paper1}.

In order to accurately extract the mass gap from the asymptotic behavior of 
Euclidean time correlation functions, various smearing techniques applied to
the hadron operator have been employed in the literature. Among them,
gaussian smearing which can be motivated from conceptually simple
considerations can be implemented in numerical simulations 
in a simple and straight forward manner.
To assess the strengths and weaknesses of the gaussian smearing, we have
carried out an extensive study that employs local-smear, smear-local and
smear-smear sink-source combinations at many values of the smearing
size parameter $s_0$. 

At each value of the smearing size,  
to extract meson observables, 
a variety of correlation functions  are measured. 
Using the pseudoscalar and the fourth component of the
axial vector densities denoted by P and A respectively, 
for the pion, PP, AA, AP and PA
correlators are used since both P and A carry quantum numbers of pion.
We have extracted mass gaps and amplitudes
(coefficients of the correlators)
for the smearing radius $s_0$ ranging from 1 to 8
in steps of 1 for all these corelation functions for all the three smearing 
combinations. At each $\kappa$, we find the optimum smearing size
$s_0$ for each correlator type; the optimum $s_0$ in general is
different for different correlators types and also varies with $\kappa$. 
In the case of pion, we do not detect any systematics in the 
optimum choice
of $s_0$ as a function of $\kappa$ as is seen from Table
\ref{piondata} except that the optimum smearing size is smaller for
AA, AP and PA than for PP. 
One should also be cautious about oversmearing. In the case of rho 
we find that $s_0$
gradually  increases as  $\kappa$ increases.

We have also carried out extensive calculations with local
operators in order to quantify
the systematic effects in the determination of masses and decay
constants.

We have obtained the pion masses from PP and also from AA
correlators. These masses are then used to determine the coefficients
for all the correlators. In each case we obtained pion decay constant and quark
masses using AA, AP and PA correlators. The values obtained using AP (PA)
correlator usually has the lowest (highest) errors. 

The general conclusion regarding the use of the different correlator
types PP, AA, AP and PA appears to be the following: whenever the axial
vector operator is at the source (i.e., AA or PA), the data 
(be it the pion mass, decay
constant or the quark mass) seems to have more noise and statistical
errors. However, despite that, these data seem to have a better
general trend overall (be it consistency check for smearing 
(Fig. \ref{ratio_coeff}),
comparison with the VWI quark mass (Fig. \ref{qmassratios}) 
or chiral trends (Figs. \ref{MsqPi_by_mQ_chiral} and
\ref{aFpi_chiral}), they also seem to
have significantly less finite volume effects.    

The quantities derived from correlators with the
psedoscalar operator (i.e., PP and AP) at the source seem to have 
the least statistical
errors, however, they seem to carry some systematic errors and
definitely show signs of stronger finite size effects at larger
$\kappa$ values.  

Although we have performed a lattice QCD investigation with fully
dynamical quarks at a host of values of $\kappa$, we do {\em not} have
nearly enough data points or small enough quark masses to have a
reliable chiral extrapolation. Let us make it very clear that although
it is one of our ultimate aims, in this paper it was not our intention
either. Given our emphasis on other aspects, e.g., study of smearing
in a very detailed manner or comparison of all pion correlators, we
still can make a few interesting observations regarding chiral trend
of our data. In both the behaviors of $aF_\pi$ and the chiral ratio 
$(a m_\pi)^2/(a m_q)$ 
versus $a m_q$, we see indication of departure from LO $\chi PT$. 
The chiral ratio has a slope downwards in the range
of $a m_q$ = 0.05 - 0.025 and our data is more consistent with NLO
$\chi PT$ than LO. Data from other works (with same parameters as
ours) done at larger volumes are
also consistent with this chiral trend including points at smaller
quark masses upto $am_q \leq 0.015$. 
The chiral ratios at $\kappa=0.158$ with $16^3 32$ lattices (from all
collaborations including ours) show significant finite size effect.

The data for $aF_\pi$  is also consistent with NLO $\chi PT$
in the range $a m_q$ = 0.05 - 0.03. The faster decrease with respect to $am_q$
at $\kappa= 0.15775$ and 0.158 is clearly due to, and in 
accordance with
$\chi PT $ finite size effects. This is again corroborated with
the simulations done at the same and larger volumes for the same 
$\kappa$ range.

We have studied the deviation of $am_q^{VWI}$ from $am_q^{AWI}$ as a 
function of $\kappa$. 
We note that $am_q^{AWI}$ has no ${\cal O}(a)$ scaling violation whereas
$am_q^{VWI}$ has some remaining ${\cal O}(a)$ scaling violations.
We have noticed small fluctuations (a few percent) around a 
normalization factor of approximately 0.78. If these deviations are
any signature, scaling violations may be small in our case.   
  
The dimensionless ratio $\frac{F_\rho}{m_\rho}$ is remarkably independent 
of $am_q$ in the range 0.07 - 0.015 and we can easily extract the constant 
value of this ratio $\approx 0.301$. Comparison with the physical value of 
this ratio (0.198) yields an estimate of $Z_V \approx 0.659$. 
The dimensionless ratio  $\frac{F_\pi}{m_\rho}$ is approximately independent
of $am_q$ in the range 0.055 - 0.03. The deviation from constancy for the 
smallest two values of $am_q$ is due to finite size effects. The constant 
value extracted for the ratio is 0.159. Comparison with the physical value of
this ratio (0.120) yields an estimate of $Z_A \approx 0.76$. It is encouraging
that the extracted values for both these renormalization constants 
(which are independent of the lattice scale) are in 
agreement with those extracted non-perturbatively on the lattice in 
Ref. \cite{deldebbio1,besirevic}.   
  
We find our initial results interesting and very encouraging.
Simulations are underway utilizing significantly larger lattices and improved
algorithms in order to further test the accessibility of chiral region to 
the naive Wilson action (both gauge and fermion sectors). It will be interesting
to compare our results obtained with unimproved Wilson fermion and gauge
action to that with improved actions, for example, Refs.
\cite{clover,domain,twisted}. 
\acknowledgments

Numerical calculations are carried out on a Cray XD1 (120 AMD
Opteron@2.2GHz) supported by the
10$^{th}$ and 11$^{th}$ Five Year Plan Projects of the Theory
Division, SINP under the
DAE, Govt. of India. This work was in part based on the MILC collaboration's
public lattice gauge theory code.
See \url{http://physics.utah.edu/~dtar/milc.html}~.
\appendix
\section{Fourier Transform Trick for Sink Smearing}
\label{ft_trick}
In this appendix, we follow the discussion in Ref. \cite{hauswirth}. 
Generically we need to calculate 
\be
\sum_{{\bf x}} \langle {\cal O}^{\rm sink}({\bf x},t) 
{\cal O}^{\rm source}({\bf x}_{0},t_0) \rangle~.
\ee
\vskip .1in
Case I (local source, local sink)
\vskip .1in
We have 
\be
\sum_{{\bf x}} 
\langle 
{\overline \psi}({\bf x},t)\Gamma \psi({\bf x},t) 
{\overline \psi}({\bf x}_{0},t_0)\Gamma \psi({\bf x}_0,t_0) 
\rangle ~ =~
\sum_{\bf x} 
\langle 
{\rm Trace} ~ (S({\bf x},t;{\bf x}_0,t_0) \Gamma S({\bf x}_0,t_0;{\bf x},t) 
\Gamma) 
\rangle~ \label{ptpt}
\ee
where $\Gamma$ denotes any of the
sixteen Dirac matrices.
The fermion propagator $S(x,y)$ is obtained by solving the equation
\be
MS=\delta
\ee
where $M$ is the Dirac-Wilson operator.  

Writing the coordinates explicitly, we have,
\be
\sum_{{\bf x}_{1}, t_{1}} M({\bf z}_{1},\tau; {\bf x}_{1},t_{1})~ 
S({\bf x}_{1},t_{1};{\bf {\bf y}}_{1},0)~=~
\delta_{{\bf z}_{1},{\bf {\bf y}}_{1}} ~ \delta_{\tau,0}~. \label{ptsource}
 \ee
For the calculation of the meson propagator, Eq. (\ref{ptpt}), 
we need to calculate only $S({\bf x},t;{\bf x}_0,t_0) $ since 
$ S({\bf x}_0,t_0;{\bf x},t)  $ is
trivially calculated from
\be
S({\bf x}_0,t_0;{\bf x},t) ~ = ~ \gamma_5 ~
S^\dagger({\bf x},t;{\bf x}_0,t_0)~ \gamma_5~.
\ee
\vskip .1in
Case II (local source, smeared sink)
\vskip .1in

\be
{\cal O}^{\rm sink} ~ &=&
  ~ \sum_{{\bf x}_{1}{\bf x}_{2}}~ \phi({\bf x}_{1}-{\bf x})~
\phi({\bf x}_{2}-{\bf x})~
{\overline \psi}({\bf x}_{1},t) \Gamma~ \psi({\bf x}_{2},t) \nonumber  \\
{\cal O}^{\rm source} ~&=&~ {\overline \psi}({\bf x}_{0},t_{0}) \Gamma ~ 
\psi({\bf x}_{0},t_{0})
~.
\ee
Now we have,
\be
\sum_{{\bf x}}\sum_{{\bf x}_{1}{\bf x}_{2}} ~\phi({\bf x}_{1}-{\bf x}) ~  \phi({\bf x}_{2} -{\bf x})~
\langle
{\rm Trace}~ S({\bf x}_{2},t; {\bf x}_0,t_0)~ \Gamma \gamma_5 
S^\dagger({\bf x}_{1},t; {\bf x}_0,t_0) \gamma_5 ~ \Gamma 
\rangle~.
\ee
Going to momentum space,
\be
\phi({\bf x}_{1}-{\bf x}_{0}) ~&=&~1/N ~ \sum_{{\bf k}_{1}}~ e^{-i {\bf k}_{1} \cdot ({\bf x}_{1}-{\bf x}_{0})}~ 
{\tilde \phi}({\bf k}_{1}),  \nonumber \\
S({\bf x}_{2},t;{\bf x}_{0},t_{0})~&=&~ 1/N ~ \sum_{{\bf k}_{3}}~ 
e^{-i {\bf k}_{3} \cdot ({\bf x}_{2}-{\bf x}_{0})}~ {\tilde S}({\bf
  k}_{3},t)~, ~{\rm etc}., 
\ee
we get,
\be
(1/N)^4~ \sum_{{\bf k}}~ {\tilde \phi}({\bf k})~{\tilde \phi}({\bf k})~
\langle {\rm Trace}~ {\tilde S}({\bf k},t) ~ \Gamma ~\gamma_5 ~ 
{\tilde S}^\dagger({\bf k},t) 
~\gamma_5~\Gamma
\rangle~. 
\ee
Need to calculate only ${\tilde S}({\bf k},t) $ and ${\tilde \phi}({\bf k}) $.

\vskip .1in
Case III (smeared source, local sink)
\vskip .1in 
\be
{\cal O}^{\rm source} ~ &=&
  ~ \sum_{{\bf x}_{1}{\bf x}_{2}}~ \phi({\bf x}_{1}-{\bf x}_{0})~\phi({\bf x}_{2}-{\bf x}_{0})~
{\overline \psi}({\bf x}_{1},t_0) \Gamma~ \psi({\bf x}_{2},t_0) \nonumber  \\
{\cal O}^{\rm sink} ~&=&~ {\overline \psi}({\bf x},t) \Gamma ~ \psi({\bf x},t)
~.
\ee
Now we have,
\be
\sum_{{\bf x}}~ \sum_{{\bf x}_{1}{\bf x}_{2}} ~ \phi({\bf x}_1-{\bf x}_0) ~\phi({\bf x}_2-{\bf x}_0)~
\langle
{\rm Trace} S({\bf x},t;{\bf x}_1,t_0) ~ \Gamma 
~\gamma_5  S^\dagger({\bf x},t;{\bf x}_2,t_0)\gamma_5 \Gamma
\rangle ~.
\ee
\be =
\sum_{\bf x} ~ 
\langle 
{\rm Trace} {\cal S}({\bf x},t;{\bf x}_0,t_0) \Gamma \gamma_5
{\cal S}^\dagger({\bf x},t;{\bf x}_0,t_0)
\gamma_5 ~ \Gamma
\rangle  
\ee
where
\be
{\cal S}({\bf x},t;{\bf x}_0,t_0)~=~ \sum_{{\bf x}_{1}}~ \phi({\bf x}_1-{\bf x}_0) S({\bf x},t;{\bf x}_{1},t_0)~.
\ee
Starting from Eq. (\ref{ptsource}),
\be
\sum_{{\bf x}_{1}}M({\bf y},t_{ y};{\bf x},t)~S({\bf x},t;{\bf x}_1,0) ~ \phi({\bf x}_{1}-{\bf x}_{0}) ~=~\sum_{{\bf x}_{1}} 
\delta({\bf y}-{\bf x}_{1}) ~\phi({\bf x}_{1}-{\bf x}_{0})~ \delta(t_y-0)~. 
\ee
i.e., 
\be
M({\bf y},t_y;{\bf x},t)~ {\cal S}({\bf x},t;{\bf x}_{0},0) ~ = ~ \phi({\bf y}-{\bf x}_{0})~ \delta(t_y)~.
\ee
Once ${\cal S}$ is calculated, ${\cal S}^\dagger$ is calculated trivially.

Note that in this case there is no need to Fourier Transform.
\vskip .1in
Case IV (smeared source, smeared sink)
\vskip .1in
In this case again it is advantageous to perform Fourier Transform and we get
\be
\sum_{\bf x} \langle  {\cal O}^{\rm sink}({\bf x},t) 
{\cal O}^{\rm source}({\bf x}_{0},t_0) \rangle ~=~ \sum_{\bf k}~ {\tilde 
\phi}({\bf k})~{\tilde \phi}({\bf k})
\langle
{\tilde {\cal S}}({\bf k},t)~\Gamma~ \gamma_5 
{\tilde {\cal S}}^\dagger({\bf k},t)
\gamma_5 ~ \Gamma 
\rangle ~.
\ee



\begin{thebibliography}{99}

\bibitem{wilson} K.~G.~Wilson, Phys.\ Rev.\ D {\bf 10}, 2445 (1974).
   K.~G.~Wilson  in {\em New Phenomena in Subnuclear Physics}, ed. A.
Zichichi
   (Plenum Press, New York), Part A, p.69 (1975).


   
\bibitem{deldebbiojhep}  L.~Del Debbio, L.~Giusti, M.~Luscher, R.~Petronzio
and
N.~Tantalo,
 JHEP {\bf 0602}, 011 (2006).

\bibitem{deldebbio1} L.~Del Debbio, L.~Giusti, M.~Luscher, R.~Petronzio and 
N.~Tantalo,
  JHEP {\bf 0702}, 056 (2007).

\bibitem{deldebbio2} L.~Del Debbio, L.~Giusti, M.~Luscher, R.~Petronzio 
and N.~Tantalo,
  JHEP {\bf 0702}, 082 (2007).

\bibitem{japo} Y.~Kuramashi,
  arXiv:0711.3938 [hep-lat].

\bibitem{qcd_paper1} Asit~K.~De, A.~Harindranath and Jyotirmoy~Maiti, 
manuscript in preparation. 

\bibitem{sommer} R.~Sommer,
  Nucl.\ Phys.\  B {\bf 411}, 839 (1994)
  [arXiv:hep-lat/9310022].

\bibitem{mcneile} Craig McNeile, 
 arXiv:0710.0985 [hep-lat].

\bibitem{parisi} G.~Parisi,
{\em Prolegomena to any future computer evaluation of the QCD mass
spectrum},
Invited talk given at Summer Inst., Progress in Gauge Field Theory, Cargese, 
France, Sep 1-15, 1983.
Published in Cargese Summer Inst. (1983). 


\bibitem{billoire} A.~Billoire, E.~Marinari and G.~Parisi,
Phys.\ Lett.\  B {\bf 162} (1985) 160.

\bibitem{aoki-stagg} S.~Aoki, M.~Fukugita, S.~Hashimoto, K.-I.~Ishikawa, 
N.~Ishizuka, Y.~Iwasaki, K.~Kanaya, T.~Kaneda, S.~Kaya, 
Y.~Kuramashi, M.~Okawa, T.~Onogi, S.~Tominaga, N.~Tsutsui, A.~Ukawa,
N.~Yamada, 
Phys.\ Rev.\  D {\bf 62}, 094501 (2000)
[arXiv:hep-lat/9912007].

\bibitem{milc3} C.~Aubin, C.~W.~Bernard, C ~DeTar, J.~Osborn, 
Steven Gottlieb. E.~B.~Gregory, D.~Toussaint, U.~M.~Heller, J.~E.~Hetrick
and R.~Sugar,
  Phys.\ Rev.\  D {\bf 70}, 114501 (2004)
  [arXiv:hep-lat/0407028].

\bibitem{alikhan} See for example,  A.~Ali Khan, S.~Aoki, G.~Boyd, 
R.~Burkhalter, S.~Ejiri, M.~Fukugita, S.~Hashimoto, N.~Ishizuka, Y.~Iwasaki, 
K.~Kanaya, T.~Kaneko, Y.~Kuramashi, T.~Manke, K.~Nagai, M.~Okawa, 
H.~P.~Shanahan, A.~Ukawa and T.~Yoshie,
  Phys.\ Rev.\  D {\bf 65}, 054505 (2002)
  [Erratum-ibid.\  D {\bf 67}, 059901 (2003)]
  [arXiv:hep-lat/0105015].

\bibitem{aoki} See for example, S.~Aoki, R.~Burkhalter, M.~Fukugita, 
S.~Hashimoto, K-I.~Ishikawa, N.~Ishizuka, Y.~Iwasaki, K.~Kanaya, T.~Kaneko, 
Y.~Kuramashi, M.~Okawa, T.~Onogi, N.~Tsutsui, A.~Ukawa, N.~Yamada and
T.~Yoshie,
Phys.~Rev.~D {\bf 68}, 054502 (2003)
[arXiv:hep-lat/0212039].


\bibitem{delo1} T.~A.~DeGrand and R.~D.~Loft,
Comput.\ Phys.\ Commun.\  {\bf 65}, 84 (1991).
Also see, T.~A.~DeGrand and R.~D.~Loft,
{\em Gaussian shell model trial wave functions for lattice QCD
spectroscopy}, 
COLO-HEP-249 (1991).



\bibitem{hauswirth} S.~Hauswirth, {\em Light hadron spectroscopy in 
quenched lattice QCD with chiral  fixed-point fermions},
arXiv:hep-lat/0204015.



\bibitem{bitar} K.~M.~Bitar, T.~Degrand, R.~Edwards, S.~Gottlieb, 
U.~M.~Heller, A.~D.~Kennedy, J.~B.~Kogut, A.~Krasnitz, W.~Liu,
M.~C.~Ogilvie, R.~L.~Renken, P.~Rossi, D.~K.~Sinclair, R.~L. Sugar, 
D.~Toussaint, and K.~C.~Wang,
Phys.\ Rev.\  D {\bf 49}, 3546 (1994)
[arXiv:hep-lat/9309011].

\bibitem{ibm} 
F.~Butler, H.~Chen, J.~Sexton, A.~Vaccarino and D.~Weingarten,
%
Nucl.\ Phys.\  B {\bf 421}, 217 (1994)
[arXiv:hep-lat/9310009];
Also see, F.~Butler, H.~Chen, J.~Sexton, A.~Vaccarino and D.~Weingarten,
Phys.\ Rev.\ Lett.\  {\bf 70}, 2849 (1993)
[arXiv:hep-lat/9212031].

\bibitem{milcw} C.~W.~Bernard,  
T.~Blum, C.~DeTar, S.~Gottlieb, U.~Heller, J.~Hetrick, C.~McNeile, 
K.~Rummukainen, B.~Sugar, D.~Toussaint, and M.~Wingate
[MILC Collaboration],
  Nucl.\ Phys.\ Proc.\ Suppl.\  {\bf 60A}, 3 (1998)
  [arXiv:hep-lat/9707014].
 


\bibitem{degrand1} T.~A.~DeGrand  [MILC Collaboration],
  [arXiv:hep-lat/9802012].

\bibitem{degrand2}T.~A.~DeGrand, A.~Hasenfratz and T.~G.~Kovacs,
  Nucl.\ Phys.\ Proc.\ Suppl.\  {\bf 73}, 903 (1999)
  [arXiv:hep-lat/9809097].


\bibitem{orthprd} B.~Orth, T.~Lippert and K.~Schilling,
  Phys.\ Rev.\  D {\bf 72}, 014503 (2005)
  [arXiv:hep-lat/0503016].


\bibitem{sokal}  Neal~Madras and Alan~D.~Sokal,
 J. Stat. Phys. {\bf 50}, 109 (1988). 

\bibitem{ape} M.~Albanese, F.~Constantini, G.~Fiorentini, F.~Flore, 
M.~P.~Lombardo, R.~Tripiccione, P.~Bacilieri, L.~Fonti, P.~Giacomelli, 
E.~Remiddi, M.~Bernaschi, N.~Cabibbo, E.~Marinari, G.~Parisi, G.~Salina,
S.~Cabasino, F.~Marzano, P.~Paolucci, S.~Petrarca, F.~Rapuano,
P.~Marchesini,  
Phys.\ Lett.\ {\bf B192}, 163 (1987).

\bibitem{cola} J.~Gasser and H.~Leutwyler,
Phys.\ Lett.\  B {\bf 184}, 83 (1987).
Also see, 
G.~Colangelo and S.~Durr,
Eur.\ Phys.\ J.\  C {\bf 33}, 543 (2004)
[arXiv:hep-lat/0311023];
G.~Colangelo, S.~Durr and C.~Haefeli,
Nucl.\ Phys.\  B {\bf 721}, 136 (2005)
[arXiv:hep-lat/0503014].



\bibitem{besirevic} D.~Be\'{c}irevi\'{c}, B.~Blossier, Ph. Boucaud, 
V.~Gim\'{e}nez, V.~Lubicz, F.~Mescia, S.~Simula and G.~Tarantino, 
  Nucl.\ Phys.\  B {\bf 734}, 138 (2006)
  [arXiv:hep-lat/0510014].

\bibitem{leut} H.~Leutwyler,
Nucl.\ Phys.\ Proc.\ Suppl.\  {\bf 94}, 108 (2001)
[arXiv:hep-ph/0011049].

\bibitem{ss} S.~R.~Sharpe and R.~L.~.~Singleton,
  Phys.\ Rev.\  D {\bf 58}, 074501 (1998)
  [arXiv:hep-lat/9804028].

\bibitem{tlqcd} Gunnar ~S.~Bali, Bram Bolder, Norbert Eicker, Thomas
Lippert, 
Boris Orth, Peer Ueberholz, Klaus Schilling, and Thorsten Struckmann, 
Phys.\ Rev.\  D {\bf 62}, 054503 (2000)
[arXiv:hep-lat/0003012].

\bibitem{clover} For example, M.~Gockeler,
Roger Horsley, Yoshifumi Nakamura, Dirk Pleiter, Paul E.~L.~ Rakow,
Gerrit Schierholz, Wolfram Schroers, Thomas Streuer, Hinnerk Stuben
and James M. Zanotti,
PoS {\bf LAT2006}, 160 (2006)
[arXiv:hep-lat/0610071].

\bibitem{domain}
For example, D.~J.~Antonio,
T.~Blum, K.~C.~Bowler, P.~.~Boyle, N.~H.~Christ, A.~D.~Cohen, M.~A.~Clark,
C.~Dawson, A.~Hart, K.~Hashimoto, T.~Izubuchi, B.~Jo\'{o}, C.~Jung,
A.~D.~Kennedy, R.~D.~Kenway, S.~Li, H.~W.~Lin, M.~F.~Lin, R.~D.~Mawhinney,
C.~M.~Maynard, J.~Noaki, S.~Ohta, S.~Sasaki, A.~Soni, R.~J.~Tweedie,
and A.~Yamaguchi 
[RBC and UKQCD Collaborations],
Phys.\ Rev.\  D {\bf 75}, 114501 (2007) [arXiv:hep-lat/0612005].

\bibitem{twisted}
For example, Ph.~Boucaud,
P.~Dimopoulos, F.~Farchioni, R.~Frezzotti, V.~Gimenez, G.~Herdoiza,
K.~Jansen, V.~Lubicz, G.~Martinelli, C.~McNeile, C,~Michael, I.~Montvay,
D.~Palao, M.~Papinutto, J.~Pickavance, G.~C.~Rossi, L.~Scorzato,
A.~Schindler, S.~Simula, C.~Urbach and U.~Wenger,
Phys.\ Lett.\  B {\bf 650}, 304 (2007)
[arXiv:hep-lat/0701012].



\end{thebibliography}
\end{document}